\documentclass[journal]{IEEEtran}
\usepackage[usenames,dvipsnames]{color}
\let\labelindent\relax
\usepackage{enumitem}
\usepackage{latexsym,psfig}
\usepackage{amsfonts}
\usepackage{amsmath,amssymb}
\usepackage{amsmath,epsfig}
\RequirePackage{color}
\usepackage{amssymb}
\usepackage{amsthm}
\usepackage{amsfonts}
\usepackage{mathrsfs}
\usepackage{algorithm}
\usepackage{algpseudocode}
\usepackage{times,lastpage,bm,xspace,subfigure,booktabs,bbm}
\usepackage{bibunits}
\usepackage{multirow}
\usepackage{url}
\usepackage{array}
\usepackage{cite}

\definecolor{DukeBlue}{RGB}{0,0,156}
\definecolor{DukeBluecmyk}{cmyk}{1,0.69,0,0.115}
\definecolor{DarkOrange}{RGB}{175,50,10}
\definecolor{BadgerRed}{RGB}{183,01,01}

\newcommand{\iid}{iid\xspace}

\newcommand{\cf}{cf.\xspace}
\newcommand{\eg}{e.g.\xspace}
\newcommand{\ie}{i.e.\xspace}
\newcommand{\h}[1]{\ensuremath{\widehat{#1}}}
\newcommand{\til}[1]{\ensuremath{\widetilde{#1}}}

\renewcommand{\cal}[1]{\ensuremath{\mathcal{#1}}}
\newcommand{\zeros}{\ensuremath{\mathbf{0}}}
\newcommand{\ones}{\ensuremath{\mathbf{1}}}

\newcommand{\complex}{\ensuremath{\mathbf{C}}}
\newcommand{\expect}{\ensuremath{\mathbf{E}}}
\newcommand{\prob}{\ensuremath{\mathbf{P}}}
\newcommand{\reals}{\ensuremath{\mathbf{R}}}
\newcommand{\kron}{\otimes} 
\newcommand{\deq}{\ensuremath{:=}}

\newcommand{\T}{\top}

\newcommand{\abs}[1]{\ensuremath{\lvert#1\rvert}}

\newcommand{\norm}[1]{\ensuremath{\lVert#1\rVert}}

\newcommand{\onenorm}[1]{\norm{#1}_1}

\newcommand{\twonorm}[1]{\norm{#1}_2}

\newcommand{\floor}[1]{\lfloor #1 \rfloor}

\DeclareMathOperator*{\argmin}{arg\,min}
\DeclareMathOperator{\diag}{diag}
\DeclareMathOperator{\RIP}{RIP}

\DeclareMathOperator{\st}{subject\ to}
\newlength{\parindentbak}
\newenvironment{aside*}
{
	\mbox{}\\\noindent\begin{minipage}{\textwidth-\parindentbak}
	\setlength{\parindent}{\parindentbak}\vline\mbox{}\indent\mbox{}%
	\begin{minipage}{\textwidth-\parindentbak}
	\setlength{\parindent}{\parindentbak}%
}
{
	\end{minipage}
	\end{minipage}\\
}

\newcommand{\hL}{h^{\mathsf{L}}}
\newcommand{\hH}{h^{\mathsf{H}}}
\newcommand{\fL}{f^{\mathsf{L}}}

\def\RIP{\text{RIP}}

\renewcommand{\vec}[1]{{#1}}
\newcommand{\mat}[1]{{\uppercase{#1}}}
\newcommand{\oper}[1]{{#1}}

\newcommand{\wh}[1]{{\widehat{#1}}}
\newcommand{\wt}[1]{{\widetilde{#1}}}

\def\FF{\mat{\mathcal{F}}}

\def\IF{\FF^{-1}}

\def\fstar{\vec{f}^{\star}}

\def\f{\vec{f}}
\def\F{\mat{F}}
\def\fest{\widehat{\f}}

\def\y{\vec{y}}

\def\A{\mat{A}}

\def\hmura{h^{\rm MURA}}
\def\hrecon{\overline{h}\! \phantom{.}^{\rm MURA}}

\def \ie{{\em i.e., }}
\def \cf{{\em cf. }}
\def \eg{{\em e.g., }}
\def\ones{\ensuremath{\mathbf{1}}}
\def\reals{\ensuremath{\mathbf{R}}}
\def\complex{\ensuremath{\mathbf{C}}}
\def\expect{\ensuremath{\mathbb{E}}}
\def\prob{\mathbf{P}}
\def\deq{\ensuremath{\triangleq}}
\def\diag{\ensuremath{\mathop{\rm diag}}}

\def\argmin{\ensuremath{\mathop{\arg \min}}}
\newtheorem{definition}{Definition}

\newtheorem{theorem}{Theorem}

\newenvironment{squishlist}
{\begin{list}{$\bullet$}
{ \setlength{\itemsep}{2pt}      \setlength{\parsep}{2pt}
  \setlength{\topsep}{0pt}       \setlength{\partopsep}{0pt}
  \setlength{\leftmargin}{1.5em} \setlength{\labelwidth}{1em}
  \setlength{\labelsep}{0.5em} } }
{\end{list}}

\begin{document}
\title{Compressive Coded Aperture Keyed Exposure Imaging with Optical Flow Reconstruction}

\author{Zachary~T.~Harmany,~\IEEEmembership{Member,~IEEE,} 
Roummel~F.~Marcia,~\IEEEmembership{Member,~IEEE,\\}
        and~Rebecca~M.~Willett,~\IEEEmembership{Senior Member,~IEEE}% <-this % stops a space
\thanks{This research is supported by DARPA Contract
    No. HR0011-04-C-0111, ONR Grant No. N00014-06-1-0610, DARPA
    Contract No. HR0011-06-C-0109, and NSF-DMS-08-11062.  
    Portions of this work were presented at the IEEE International Conference on Acoustic, Speech, and Signal Processing, March 2008, and at the SPIE Electronic Imaging Conference, January 2009,
    and SPIE Photonics Europe, April 2010.
}% <-this % stops a space
\thanks{Z.T.~Harmany is with the Department of Electrical and Computer Engineering, The University of Wisconsin-Madison, Madison, WI 53706 USA (e-mail: harmany@wisc.edu).}
\thanks{R.M.~Willett is with the Department of Electrical and Computer Engineering, Duke University, Durham, NC 27708 USA (e-mail: willett@duke.edu).}
\thanks{R.F.~Marcia is with the School of Natural Sciences, University of California, Merced, CA 95343 USA (e-mail: rmarcia@ucmerced.edu).
}%
}
%\thanks{Manuscript received April 19, 2005; revised January 11, 2007.}}

% and the end of the author line. \ie if you had this:

\markboth{}{}



\maketitle

% ==================================================
% = Abstract                                       =
% ==================================================
\begin{abstract}
%\boldmath

%Optical systems which measure arbitrarily-structured independent random projections of a scene according to compressed sensing (CS) theory face a myriad of practical challenges related to the size of the physical platform, photon efficiency, the need for high temporal resolution, and fast reconstruction in video settings. 

This paper describes a coded aperture and keyed exposure approach to compressive video measurement which admits a small physical platform, high photon efficiency, high temporal resolution, and fast reconstruction algorithms.
The proposed projections satisfy the Restricted Isometry Property (RIP), and hence compressed sensing theory provides theoretical guarantees on the video reconstruction quality. Moreover, the projections can be easily implemented using existing optical elements such as spatial light modulators (SLMs). We extend these coded mask designs to novel dual-scale masks (DSMs) which enable the recovery of a coarse-resolution estimate of the scene with negligible computational cost. We develop fast numerical algorithms which utilize both temporal correlations and optical flow in the video sequence as well as the innovative structure of the projections. Our numerical experiments demonstrate the efficacy of the proposed approach on short-wave infrared data.
\end{abstract}


% \fi
\IEEEpeerreviewmaketitle


% ==================================================
% = Introduction                                   =
% ==================================================
\section{Introduction}

\IEEEPARstart{T}{he} theory of compressed sensing (CS) suggests that we can collect high-resolution imagery with relatively few photodetectors or a small focal plane array (FPA) when the scene is sparse or compressible in some dictionary or basis and the measurements are chosen appropriately \cite{CS:candes1,CS:donoho}. There has been significant recent interest in building imaging systems in a variety of contexts to exploit CS (\cf \cite{Shankar:08, Coskun:10, Potter:10, Gu:08, SparseDNA, CS_DNAMicroarrays, LustigMRI, GPR, confocal, CSastronomy}). By designing optical sensors to collect measurements of a scene according to CS theory, we can use sophisticated computational methods to infer critical scene structure and content. One particularly famous example is the Single Pixel Camera \cite{riceCamera}, which collects sequential projections of the scene. While these measurements are supported by the CS literature, there are several practical challenges associated with the tradeoff between temporal resolution and physical system footprint. In this paper we describe an alternative approach to designing a low frame-rate {\em snapshot} CS camera which naturally parallelizes the compressive data acquisition. Our approach is based on two imaging techniques called coded apertures and keyed exposures, which we explain next.

Coded apertures \cite{ables, dicke} have a long history in low-light astronomical applications. Coded apertures were first developed to increase the amount of light hitting a detector in an optical system without sacrificing resolution (by, say, increasing the diameter of an opening in a pinhole camera). The basic idea is to use a mask, \ie an opaque rectangular plate with a specified pattern of openings, that allows significantly brighter observations with higher signal-to-noise ratio than those from conventional pinhole cameras \cite{ables, dicke}.  These masks encode the image before detection, inducing a more complicated point spread function than that associated with a pinhole aperture. The original scene is then recovered from the encoded observations in post-processing using an appropriate reconstruction algorithm which exploits the mask pattern.  These multiplexing techniques are particularly popular in astronomical \cite{caroli, skinner} and medical \cite{accorsi, gindi, meikle} applications because of their efficacy at wavelengths where lenses cannot be used, but recent work has also demonstrated their utility for collecting both high resolution images and object depth information simultaneously \cite{freeman}.

Keyed exposure imaging (also called coded exposure \cite{codedExposure}, flutter shutter \cite{flutterShutter}, or coded strobing \cite{codedStrobing}) was initially developed to facilitate motion deblurring in video using a relatively low frame rate. In some cases motion has been inferred from a single keyed exposure snapshot. The basic idea is that the camera sensor continuously collects light while the shutter is rapidly opened and closed; the shutter movement effectively modulates the motion blur point spread function, and with well-chosen shutter movement patterns it becomes possible to deblur moving objects. Similar effects can be achieved using a strobe light instead of moving a shutter. 

Despite the utility of the above methods in specific settings, they both face some limitations. The design of conventional coded apertures does not account for the inherent structure and compressibility of natural scenes, nor the potential for nonlinear reconstruction algorithms. Likewise, existing keyed exposure methods focus on direct (uncoded) measurements of the spatial content of the scene and have limited reconstruction capabilities, as we detail below. The Coded Aperture Keyed Exposure (CAKE) sensing paradigm we describe in this paper is designed to allow nonlinear high-resolution video reconstruction from relatively few measurements in more general settings. Recent related work \cite{llull2013coded-aperture-compressive-temporal-imaging} following our early technical report \cite{cake_arxiv} describes a physical platform for combining coded apertures with coded exposures. We provide theoretical support for this framework as well as detail novel dual-scale mask generation schemes that can be used to compute a coarse-resolution estimate of the scene with negligible computational cost; this coarse estimation can be followed by reconstruction algorithms which use optical flow models of video sequences.

\subsection{Problem Formulation}
\label{sec:probform}

We consider the problem of reconstructing an $N$-frame video sequence $\fstar$, where each frame is an $n_1 \times n_2$ two-dimensional image denoted $\fstar_t$. Using standard vector representation, we have that $\fstar_t \in \reals^n$ for $t = 1,\ldots,N$ where $n \deq n_1 n_2$ is the total number of pixels. As a result, the vector representation of the video sequence is $\fstar = [(\fstar_1)^\T, \ldots, (\fstar_N)^\T]^\T \in \reals^{nN}$.

The observations $y$ of $\fstar$ are also acquired as a video sequence. We do not assume that the observations are acquired at the same rate at which we will ultimately reconstruct $\fstar$. In general, we assume $y$ is an $M$-frame video sequence, with each frame $y_k$ of size $m_1 \times m_2$. Similarly to $\fstar$, we have $y_k \in \reals^m$ for $k = 1,\ldots,M$, where $m \deq m_1 m_2$, therefore $y = [y_1^\T, \ldots, y_M^\T]^\T \in \reals^{mM}$.

We observe $\fstar$ via a spatio-temporal \emph{sensing matrix} $A \in \reals^{mM \times nN}$ which linearly projects the spatio-temporal scene onto an $mM$-dimensional set of observations:
\begin{equation*} \label{eq:obs}
	\vec{y} = A\fstar + w,
\end{equation*}
where $w \in \reals^{mM}$ is noise associated with the physics of the sensor. 
%This acquisition model specializes to a snapshot camera when $N = M = 1$. 

CS optical imaging systems must be designed to meet several competing objectives:
\begin{squishlist}
\item The sensing matrix $A$ must satisfy some necessary criterion (such as the RIP, defined below) which provides theoretical guarantees on the accuracy with which we can estimate $\fstar$ from $y$.
\item The total number of measurements, $mM$, must be lower than the total number of pixels to be reconstructed, $nN$. This is achievable via compressive spatial acquisition ($m < n$), frame rate reduction ($M < N$), or simultaneous spatio-temporal compression.
\item The measurements $y$ must be {\em causally} related to the temporal scene $\fstar$, which restricts the structure of the projections $A$.
\item The optical measurements modeled by $A$ must be implementable in a way that results in a smaller, cheaper, more robust, or lower power system.
\item The sensing matrix structure must facilitate fast reconstruction algorithms.
\end{squishlist}
{\em This paper demonstrates that compressive Coded Aperture Keyed Exposure  systems achieve all these objectives.}

\subsection{Contributions} 

The primary contribution of this paper is the design and theoretical characterization of compressive Coded Aperture Keyed Exposure (CAKE) sensing. We describe how keyed exposure ideas can be used in conjunction with coded apertures to increase both the spatial and temporal resolution of video from relatively few measurements. We prove hitherto unknown theoretical properties of such systems and demonstrate their efficacy in several simulations. We also detail a new mask design method that enables, with trivial computational cost, a low-resolution image of the scene from which we estimate an optical flow vector field for use in a computationally efficient optical flow-based reconstruction algorithm. In addition, we discuss several important algorithmic aspects of our approach, as well as detailing considerations for hardware implementations of the proposed approach. This paper builds substantially upon earlier preliminary studies by the authors \cite{marciaICASSP,MarciaEUSIPCO,Marcia:09,MarciaHW_SPIE2010} and related independent work by Romberg \cite{RombergToeplitz}.

\subsection{Organization of the Paper} 

The paper is organized as follows. Section~\ref{sec:background} introduces relevant background material; Section~\ref{sec:arch} describes conventional coded aperture imaging techniques and Section~\ref{sec:codedexposure} describes keyed exposure techniques currently in the literature. We describe the compressive sensing problem and formulate it mathematically in Section~\ref{sec:CS}. We describe our approach to video compressed sensing using the CAKE paradigm in Section~\ref{sec:cake} including theoretical analysis of the method. We then describe the design of novel dual-scale mask patterns in Section~\ref{sec:dsm}. Section~\ref{sec:algo} details how we perform the video reconstruction from the CAKE measurements which we validate with numerical experiments in Section~\ref{sec:experiments}. We conclude with a discussion of the implementation details and tradeoffs in practical optical systems in Section~\ref{sec:hardware}.

% ==================================================
% = Background                                     =
% ==================================================
\section{Background}
\label{sec:background}

Prior to detailing our main contributions, we first review pertinent background material. This review touches upon the development of coded aperture imaging, coded exposure photography, and a brief review of salient concepts in compressed sensing theory. 

\subsection{Coded Aperture Imaging} 
\label{sec:arch}

Seminal work in coded aperture imaging includes the development of masks based on Hadamard transform optics \cite{Sloane:76} and pseudorandom phase masks \cite{Ashok:07}. Modified Uniformly Redundant Arrays (MURAs) \cite{mura} are generally accepted as optimal mask patterns for coded aperture imaging.  These mask patterns (which we denote by $\hmura$) are binary, square patterns, whose \emph{grid size matches the spatial resolution of the photo-detector} and hence are length $m$. Each mask pattern is specifically designed to have a complementary pattern $\hrecon$ such that the two-dimensional convolution $\hmura * \hrecon$ is a single peak with flat side-lobes (\ie a Kronecker $\delta$ function).

In practice, the resolution of a detector array dictates the properties of the mask pattern and hence the resolution at which $\fstar$ can be reconstructed. We model this effect as $\fstar$ being downsampled to the resolution of the detector array and then convolved with the mask pattern $\hmura$, which has the same resolution as the FPA and the downsampled $\fstar$, \ie
\begin{equation*} \label{eq:mura}
	y = ( D \fstar) * \hmura + w,
\end{equation*}
where $*$ denotes circular convolution in two dimensions, $w$ corresponds to noise associated with the physics of the sensor, and $D \fstar$ is the \emph{integration} downsampling of the scene, which consists of partitioning $\fstar$ into uniformly sized $d_1 \times d_2$ blocks, where $d_i \deq n_i/m_i$ for $i = 1,2$, and measuring the total intensity in each block.

Because of the construction of $\hmura$ and $\hrecon$, $D\fstar$
can be reconstructed using
$\fest = y * \hrecon$.
However, the resulting resolution is often lower than what is necessary to capture some of the desired details in the image. Clearly, the estimates from MURA reconstruction are limited by the spatial resolution of the photo-detector.  Thus, high resolution reconstructions cannot generally be obtained from low-resolution MURA-coded observations. 

It can be shown that this mask design and reconstruction result in minimal reconstruction errors \emph{at the FPA resolution} and \emph{subject to the constraint that linear, convolution-based reconstruction methods would be used}.  However, when the scene of interest is sparse or compressible, and nonlinear sparse reconstruction methods may be employed, then CS ideas can be used to design coded apertures.

\subsection{Coded (Keyed) Exposure Imaging}
\label{sec:codedexposure}

Coded (or keyed) exposures  were developed recently in the computational photography community. Initial work in this area was focused on engineering the temporal component of a motion blur point spread function by rapidly opening and closing the shutter during a single exposure or a small number of exposures at a low frame rate \cite{codedExposure, flutterShutter}. That is,
\begin{equation*}
y = A^{\rm KE} \fstar + w = \sum_{t \in T} \fstar_{t} + w,
\end{equation*}
where the keyed exposure (KE) measurement matrix $A^{\rm KE}$ selects the subset of frames
during which the shutter is open.  We refer to this subset as the 
exposure code $T \subseteq \{1,\ldots,N\}$.

If an object is moving during image acquisition, then a static shutter would induce a typical motion blur, making the moving object difficult to resolve with standard deblurring methods. However, by ``fluttering'' the shutter during the exposure using carefully designed patterns, the induced motion blur can be made invertible and moving objects can be accurately reconstructed. Instead of a moving shutter, more recent work uses a strobe light to produce a similar effect \cite{codedStrobing}.

While this novel approach to video acquisition can produce very accurate deblurred images of moving objects, there is significant overhead associated with the reconstruction process. To see why, note that every object moving with a different velocity or trajectory will produce a different motion blur. This means that (a) any stationary background must be removed during preprocessing and (b) multiple moving objects must be separated and processed individually. 

More recently, it was shown that these challenges could be sidestepped when the video is temporally periodic (\eg consider a video of an electronic toothbrush spinning) \cite{codedStrobing}. The periodic assumption amounts to a sparse temporal Fourier transform of the video, and this approach, called coded strobing, is a compressive acquisition in the temporal domain. As a result, the authors were able to leverage ideas from compressed sensing to achieve high-quality video reconstruction.

The assumption of a periodic video makes it possible to apply much more general reconstruction algorithms that do not require background subtraction or separating different moving components. However, it is a very strong assumption, which places some limits in its applicability to real-world settings. The approach described in this paper has similar performance guarantees but operates on much more general video sequences.

\subsection{Compressed Sensing}
\label{sec:CS}

In this section we briefly define the Restricted Isometry Property (RIP) and explain its significance to reconstruction performance.  In subsequent sections, we demonstrate our primary theoretical contribution, which is to prove the RIP for compressive coded aperture and keyed exposure systems.

\begin{definition}[Restricted Isometry Property (RIP) \cite{RIP}]
A matrix $A$ satisfies the RIP of order $s$ if there exists a constant $\delta_s \in (0,1)$ for which
\begin{equation*} \label{eq:RIP}
  (1 - \delta_{s}) \| f \|_2^2 \le \|A f \|_2^2 \le
  (1 + \delta_{s}) \| f \|_2^2.
\end{equation*}
holds for all $s$-sparse $f \in \reals^n$. If this property holds, we say $A$ is $\RIP(s,\delta_{s}).$
\end{definition}
Matrices which satisfy the RIP  are called CS matrices; and when combined with sparse recovery algorithms, they are guaranteed to yield accurate estimates of the underlying function $\fstar$:
\begin{theorem}[Sparse Recovery with RIP Matrices \cite{candesTutorial,Candes:06c}]
\label{thm:recovery2}
Let $A$ be a matrix satisfying $\mbox{RIP}(2s,\delta_{2s})$ with $\delta_{2s} < \sqrt{2}-1$, and let $y = Af + w$ be a vector of noisy observations of any signal $f \in \reals^{n}$, where the $w$ is a noise or error term with $\|w\|_{2}\leq \epsilon$ for some $\epsilon > 0$. Let $f_s$ be the best $s$-sparse approximation of $f$; that is, $f_s$ is the approximation obtained by keeping the $s$ largest entries of $f$ and setting the others to zero. Then the estimate
\begin{equation} \label{eq:constrained}
\begin{aligned}
\fest \ = \ &\argmin_{f \in \reals^n} 	& &\|f\|_{1}\\
		 &\st						& &\|y-Af\|_{2} \leq \epsilon,
\end{aligned}
\end{equation}
obeys
\begin{equation*}
\| f - \fest \|_{2} \leq C_{1,s} \epsilon + C_{1,s} \frac{\|f-f_{s}\|_{1}}{\sqrt{s}},
\end{equation*}
where $C_{1,s}$ and $C_{2,s}$ are constants which depend on $s$ but not on $n$ or $m$.
\end{theorem}
\noindent Note that the reconstruction \eqref{eq:constrained} in Theorem~\ref{thm:recovery2} is equivalent to
\begin{equation} \label{eq:l1regularized}
\fest = \argmin_{f \in \reals^n} \tfrac{1}{2}\|y -Af\|_2^2 + \tau \| f\|_1 
\end{equation}
where $\tau>0$, which depends on $\epsilon$, can be viewed as a regularization parameter.

% ==================================================
% = Coded Aperture Keyed Exposure Sensing          =
% ==================================================
\section{Coded Aperture Keyed Exposure Sensing}
\label{sec:cake}

Here we detail our compressive video acquisition method that combines coded apertures and keyed exposures to address all the competing challenges detailed in Sec.~\ref{sec:probform}. In our CAKE imaging method, each observed frame $y_k$ is given by an exposure of $B \deq N/M$ high-rate coded observations in the time interval $T_k \deq \{(k-1)B+1,\ldots, kB\}$:
\begin{equation}
\label{eq:cake}
y_k = \sum_{t \in T_k} A_t \fstar_t + w_k,
\end{equation}
for $k = 1,\ldots,M$ and $w_k$ is a zero-mean additive noise. Without the exposure, the action of each $A_t$ would be to collect a compressive snapshot measurement of the scene at time $t$. Each of these snapshots utilizes a high-resolution coded aperture and can be modeled mathematically as
\begin{equation}
\label{eq:cakesingle}
A_t \fstar_t = S(\fstar_t * h_t);
\end{equation}
where $h_t \in \reals^n$ is a coding mask, and $S \in \{0,1\}^{m \times n}$ is a subsampling operator (detailed below). The subsampling operator effectively models the action of a low-resolution FPA or CCD array. Our theory allows $S$ to be a structured nonrandom linear operator, and hence we can rewrite the above as
\begin{equation}
\label{eq:cakeconv}
y_k = S \left[ \sum_{t \in T_k} (\fstar_t * h_t) \right].
\end{equation}
To implement this sensing paradigm we simply modulate the coded aperture mask over the $B$-frame exposure time. 
%Because of this, one can think of our system as performing a coded aperture acquisition \emph{for each coded aperture element}.

It should be stressed that the coding masks ($\{ h_t\}_{t=1}^N$) are of the size and resolution at which $\fstar$ will be reconstructed; this is in contradistinction to the MURA system, in which $\hmura$ is at the size and resolution of the FPA. Thus in \eqref{eq:cakesingle}, we model the measurements as the scene being convolved with the coded mask and \emph{then} downsampled. 

While the sensing strategy described in \eqref{eq:cake} is straightforward to implement, the additional structure imposed on the sensing matrix complicates the performance analysis of the system from a compressive sensing perspective. To highlight this, we now describe the structure of these sensing operations.

The two-dimensional convolution of $h_t$ with a frame $\fstar_t$ as in (\ref{eq:cakeconv}) can be represented as the application of the Fourier transform to $\fstar_t$ and $h_t$, followed by element-wise multiplication and application of the inverse Fourier transform. In matrix notation, this series of linear operations can be expressed as
\begin{equation*} \label{eq:R}
	(\fstar_t * h_t)  = \IF \diag(\FF h_t) \FF \fstar = H_t \fstar_t,
\end{equation*}
where $\FF$ is the two-dimensional Fourier transform matrix, and $\diag(\FF h_t)$ is a diagonal matrix whose elements correspond to the transfer function, which is the Fourier transform of $h_t$. The matrix product $H_t = \IF \diag(\FF h_t) \FF \in \reals^{n \times n}$ is block-circulant with circulant block (BCCB) matrix. In matrix notation, a BCCB matrix $H$ is consists of $n_2 \times n_2$ blocks,
\begin{equation*}
	H = 
	\begin{bmatrix}
		H_{1} 		& H_{n_2}		& \cdots & H_3 		& H_2 \\
		H_{2} 		& H_1 			& \cdots & H_4 		& H_3 \\
		\vdots 		& \vdots 		& \ddots & \ddots 	& \vdots \\
		H_{n_2} 	& H_{n_2-1} 	& \cdots & H_2  	& H_1
	\end{bmatrix},
\label{eq:blockcirc}
\end{equation*}
where each $H_j \in \reals^{n_1 \times n_1}$ is circulant; \ie
$H_j$ is of the form
\begin{equation*}
	H_j =
	\begin{bmatrix}
		h_{j,1} 		& h_{j,n_1} 	& \cdots & h_{j,3}	& h_{j,2} \\
		h_{j,2} 		& h_{j,1} 		& \cdots & h_{j,4} 	& h_{j,3} \\
		\vdots 		    & \vdots 		& \ddots & \ddots 	& \vdots  \\
		h_{j,n_1} 	    & h_{j,n_1-1} 	& \cdots & h_{j,2} 	& h_{j,1} 
	\end{bmatrix}.
\end{equation*}
This structure is a direct result of the fact that the $k$-point one-dimensional Fourier transform $F_k$ diagonalizes any $k \times k$ circulant matrix (such as $H_j$ with $k=n_1$) and so ${\FF} \equiv F_{n_2} \otimes F_{n_1}$ diagonalizes block-circulant matrices (such as $H$). Here $\otimes$ denotes the Kronecker matrix product.

% \begin{figure}
% \begin{center}
% \includegraphics[width=\columnwidth]{figures/Circ_unsym.pdf}
% \caption{The $n \times n$ matrix $\IF \diag(\FF h) \FF$ is block-circulant with $n_2$ blocks in each row and column. Each block is $n_1 \times n_1$ and is circulant.}
% \label{fig:circulant} 
% \end{center}
% \end{figure} 

For each frame, a compressed sensing matrix can be formed from the matrix $H$ by retaining only a subset of the rows of $H_t$. This is equivalent to restricting the number of measurements collected of the vector $H_t \fstar_t$. Here we simply subsample $H_t \fstar_t$ by applying a point-wise subsampling matrix $S$ which retains only one measurement per uniformly sized $d_1 \times d_2$ block; \ie we subsample by $d_1$ in the first coordinate and $d_2$ in the second coordinate so that the result is an $m_1 \times m_2$ image with $m_i = n_i/d_i, i = 1,2$. 
%For compactness of notation we assume that $d_i$ evenly divides $n_i$ for $i=1,2$. 
In matrix form $S$ can be thought of as retaining a certain number of rows of the identity matrix. Because of the structure and deterministic nature of this type of downsampling, it is often more straightforward to realize in practical imaging hardware (see Sec.~\ref{sec:hardware}).

\begin{figure}[t]
   \centering
   \includegraphics[width=\columnwidth]{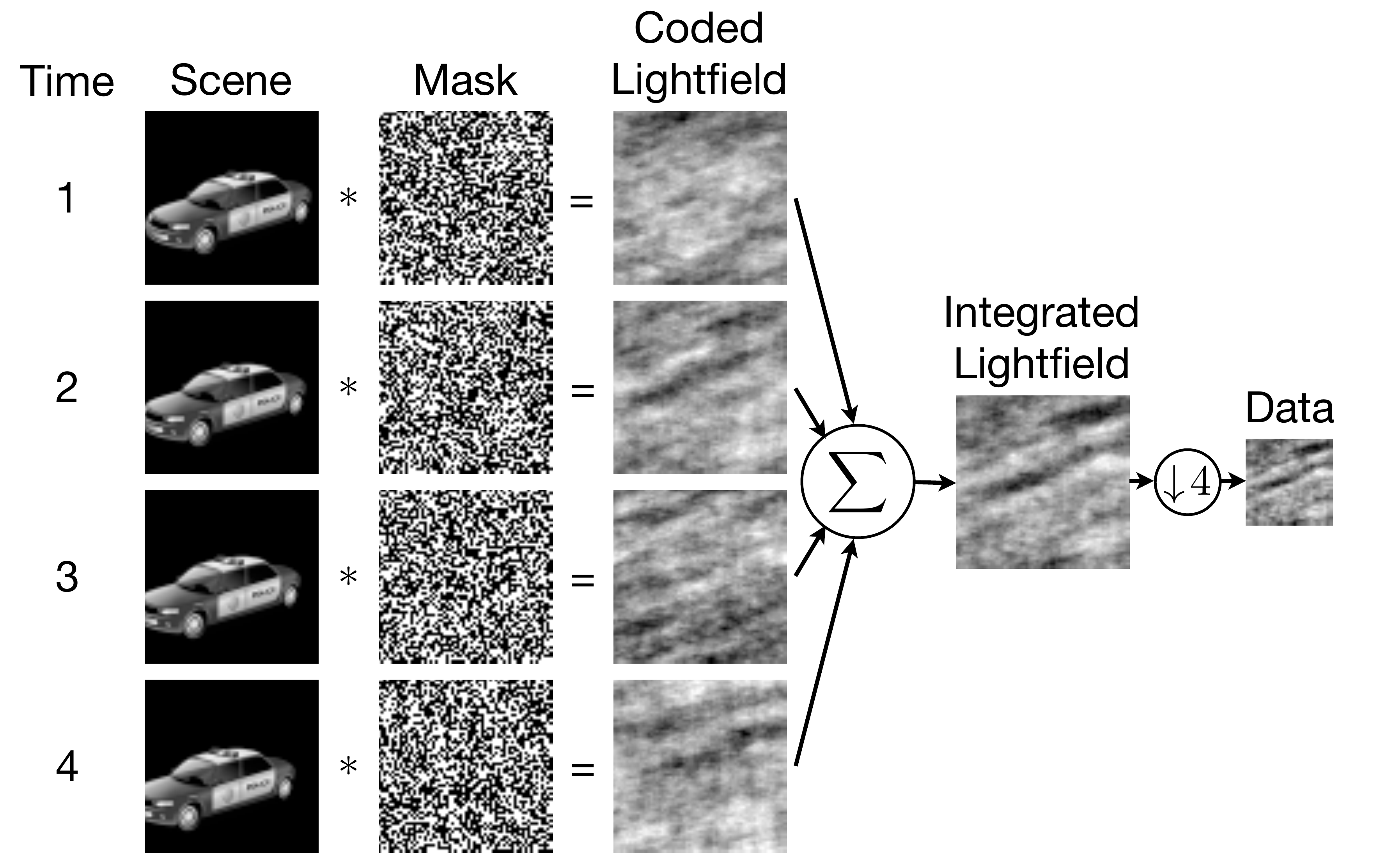}
   \caption{Illustration of a single measured frame using CAKE sensing.}
   \label{fig:CAKEExample}
\end{figure}

The resulting per-frame sensing matrix $A_t$ is then given by
\begin{equation*}
A_t = S H_t = S \IF \diag(\FF h_t) \FF.
\end{equation*}
For our CAKE sensing strategy in \eqref{eq:cake}, the sensing matrix for each block of $B$ frames is formed by concatenating sensing matrices of the above type. For the $k$th low-rate observation, we have a corresponding sensing matrix
\begin{align*}
A_k &= \begin{bmatrix} A_{(k-1)B+1} & \cdots & A_{kB} \end{bmatrix} \\
&= S \begin{bmatrix} H_{(k-1)B+1} & \cdots & H_{kB} \end{bmatrix}.
\end{align*}
We illustrate this sensing procedure in Figure~\ref{fig:CAKEExample}.

Since our sensing strategy is independent across each low-resolution frame, and also for simplicity of presentation, we only consider the recovery of a length $B$ block of frames from a single snapshot image in our theoretical analysis. 

\begin{theorem}[Coded Aperture Keyed Exposure Sensing]
\label{thm:cake} Let $A = [A_1 \cdots A_B]$ be an $m \times nB$ sensing matrix for the CAKE system where the coded aperture pattern for each $A_t$ is drawn \iid from a suitable probability distribution. Then there exists absolute constants $c_1, c_2 > 0$ such that for any
\begin{equation*}
m \ge c_1 \delta^{-2} s^2 \log(nB),
\end{equation*}
$A$ is $\RIP(s,\delta_s)$ with $\delta_s \le \delta$ with probability exceeding
\begin{equation*}
	1 - 2n^2B^2 e^{-c_2 m/s^2}.
\end{equation*}
\end{theorem}
Note here that $s$ denotes the sparsity of $B$ frames of the video sequence, rather than simply the sparsity of an individual frame; that is, if each frame had sparsity $s'$, then na\"ively, without accounting for temporal correlations, we would have $s=Bs'$. We present the proof of Theorem~\ref{thm:cake} in Appendix~\ref{sec:proof} for the particular case where we select
\begin{equation*} \label{eq:rademacher}
[h_t]_k = 
\begin{cases} 
\phantom{-}\sqrt{d/n} & \text{ with probability } 1/2, \\
         - \sqrt{d/n} & \text{ with probability } 1/2, 
\end{cases}
\end{equation*}
\iid for $k = 1,\ldots,n$ and $t = 1,\ldots,N$. This binary-valued distribution was selected to model coded apertures with two states -- open and closed -- per mask element. It is straightforward to extend the result to other mask generating distributions, such as uniform or Gaussian distributed entries. We discuss the issue of implementing these masks in Section~\ref{sec:hardware}.

Recent work by Bajwa et al.~\cite{waheed, haupttoeplitz}, showed that subsampled rows from random circulant matrices (and Toeplitz matrices, in general) are sufficient to recover sparse $\f^{\star}$ from $\y$ with high probability.  In particular, they showed that when $m\geq C \delta^{-2} s^2 \log n$ for some constant $C > 0$, a Toeplitz matrix whose first row contains elements drawn independently from a Gaussian distribution are $\RIP(s,\delta_s)$ with $\delta_s \le \delta$ with high probability, with extensions to other generating distributions. Drawing $h$ \iid from any zero-mean sub-Gaussian distribution with variance $d/n$ is sufficient to establish RIP; if one is willing to pay additional logarithmic factors \cite{krahmermendelson13suprema} shows RIP can be established if $m \ge C \delta^{-2} s (\log s)^2 (\log n)^2$ for some constant $C$. 

It is of interest to note that since the CAKE sensing matrix is a concatenation of downsampled Toeplitz matricies, this sensing strategy has clear connections to \cite{WaheedMud} where they consider concatenations of Toeplitz matricies as a sensing matrix for performing multiuser detection in wireless sensor networks. The important conceptual link is that their sensing matrix is used to determine a sparse set of simultaneously active users, where in our system, we are using it to infer a sequence of sparse frames.

\section{Dual-scale masks}
\label{sec:dsm}

While the \iid random coded aperture mask patterns are supported theoretically by Theorem~\ref{thm:cake}, using a generative distribution with more structure allows for mask designs with several practical benefits. Here we describe a new method of coded aperture mask design which we call Dual-Scale Masks (DSMs). This approach allows us to obtain a coarse low-rate low-resolution estimate from which we can often accurately estimate the optical flow motion field. We describe the algorithms used for optical flow estimation as well as how these optical flow estimates are used in the final reconstruction in Sec.~\ref{sec:opticalflow}. As these mask designs leverage elements of Romberg's work \cite{RombergToeplitz}, we first review important components of this work prior to detailing our dual-scale mask designs.

\subsection{Phase-Shifting Mask Designs}
\label{sec:phase}

We note that Romberg \cite{RombergToeplitz} conducted related work concurrent and independent of our initial investigations \cite{marciaICASSP}. While the random convolution in Romberg's work follows a similar structure, the convolution pattern is generated randomly in the frequency domain. More specifically, the entries of the transfer function correspond to random \emph{phase shifts} (with some constraints to keep the resulting observation matrix real-valued). For simplicity of presentation, we describe the generating distribution for a one-dimensional convolution for even-length masks. First we generate $\sigma \in \complex^n$ such that
\begin{equation}
    [\sigma]_l 
    = 
    \begin{cases} \pm 1 \text{ with equal probability}	& \text{if $l=1$}, \\
        e^{i \phi} \text{ with $\phi \sim \mathcal{U}(0, 2\pi)$} & \text{if $2 \le l \le n/2$}, \\
		\pm 1 \text{ with equal probability}	& \text{if $l=n/2+1$} \\
		[\sigma]_{n-l+2}^* & \text{if $n/2 + 2 \le l \le n$}.
    \end{cases}
\label{eq:psmdist}
\end{equation}
Here $\mathcal{U}(0, 2\pi)$ denotes the uniform distribution over $[0, 2 \pi]$. Then the real-valued convolution kernel is then given by $h = F^{-1} \sigma$. This result can be extended easily for two-dimensional convolutions. We call coded aperture mask patterns $h$ generated by this procedure \emph{phase-shift masks}. 

To form a compressed sensing matrix from this random convolution, \cite{RombergToeplitz} considers two different downsampling strategies: sampling at random locations and random demodulation. The first method entails selecting a random subset $\Omega \subset \{1,\ldots,n\}$ of indices. We form a downsampling matrix $S_\Omega$ by retaining only the rows of the identity matrix indexed by $\Omega$, hence the resulting measurement matrix is given by
\begin{equation*}
A= S_\Omega \IF \diag(\sigma) \FF,
\label{eq:psmsubsample}
\end{equation*}
where $\sigma$ is generated according to the two-dimensional analog of \eqref{eq:psmdist}. The random demodulation method multiplies the result of the random convolution by a random sign sequence $s \in \{-1,1\}^n$, such that $s_k = \pm 1$ with equal probability for all $k = 1, \ldots, n$, then performs an integration downsampling of the result. Therefore in this case the measurement matrix is
\begin{equation*}
\A = D \diag(s) \IF \diag(\sigma) \FF.
\label{eq:psmdemodulate}
\end{equation*}
It can be shown that both of these strategies yield RIP-satisfying matrices with high probability. 

\begin{theorem}[Fourier-Domain CCA Sensing~\cite{RombergToeplitz}]
Let $\A$ be an $m \times n$ sensing matrix resulting from random convolution with phase shifts followed by either random subsampling or random demodulation, and let $W$ denote an arbitrary orthonormal basis. Then there exists a constant $c_3>0$ such that if \begin{equation*}
m \ge c_3 \delta^{-2} \min(s \log^6 n, s^2 \log^2 n),
\end{equation*}
$\A W$ is $\RIP(s,\delta_{s})$ with $\delta_{s} \le \delta$ with probability exceeding $1 - O(n^{-1})$.
\end{theorem}

In certain regimes, these theoretical results are stronger than those in Theorem~\ref{thm:cake} specialized to $B=1$. The main strength is that these results allow sparsity in an arbitrary orthonormal basis. However, there are important differences between the observation models which have a significant impact on the feasibility and ease of hardware implementation. We elaborate on this in Section~\ref{sec:hardware}. 

\subsection{Dual-Scale Mask Designs}
\label{sec:dsmdetails}

As the name suggests, our DSMs utilize two mask patterns; one pattern is related to the coarse low-rate low-resolution measurement scale, and the other at the fine high-rate high-resolution reconstruction scale. We differentiate the two patterns using superscripts: $\hL_k$ for the coarse-scale masks, and $\hH_t$ for the fine-scale masks. We leverage the uniform phase generation strategy in Section~\ref{sec:phase} for our low-resolution pattern, and generate our high-resolution pattern directly in the spatial domain. The reason being that the random phase patterns yield a unitary transform which is easily inverted to acquire a low-resolution estimate. Generating the high-resolution pattern in the spatial domain easily allows us to ensure that each low-resolution block in the pattern has zero mean. While this may be possible with a suitable normalization of the phase-shift patterns, some of the desirable properties of these patterns will be lost.

\begin{figure}[t]
   \centering
   \includegraphics[width = 0.75 \columnwidth]{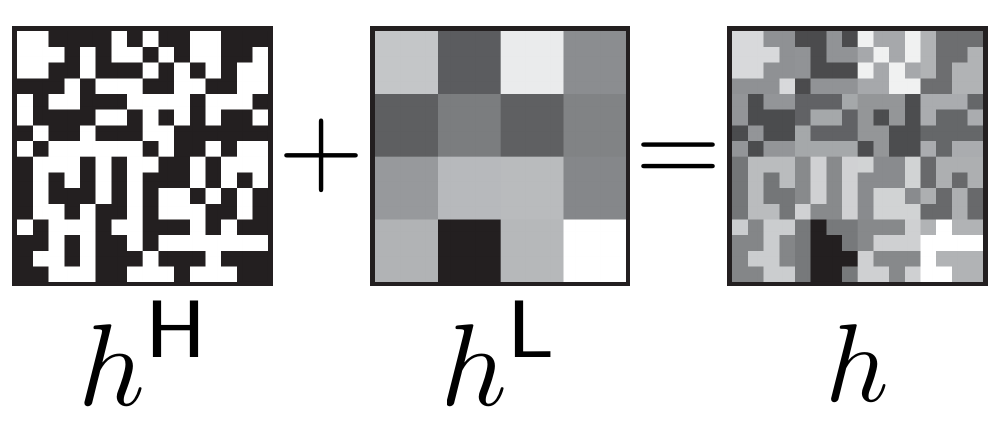}
   \caption{Example dual-scale mask pattern. 
   Here $n_1 = n_2 = 16$ ($n = 256$), $d_1 = d_2 = 4$ ($d=16$), 
   and thus $m_1 = m_2 = 4$ ($m=16$).}
   \label{fig:DSMCCA}
\end{figure}

Our procedure is as follows. For $k = 1,\ldots,M$, generate a length-$m$ random phase sequence $\sigma_k \in \complex^m$ such that $\abs{[\sigma_k]_l} = 1$, $l = 1, \ldots, m$ and $\sigma_k$ is conjugate symmetric (\ie according to the two-dimensional analog of \eqref{eq:psmdist} with $n=m$). We then let $\hL_k = D^\T \FF^{-1} \sigma_k$ where $D$ is integration downsampling and $\FF$ is the 2D DFT matrix. For $t = 1,\ldots,N$, generate an $\hH_t$ as follows. First we partition each mask into $d_1 \times d_2$ blocks, then on each block we select a binary-valued vector in $\{-\sqrt{d/n}, \sqrt{d/n}\}^d$ uniformly at random from binary vectors which sum to zero and assign these values to this portion of the block. Said differently, on each block we randomly select half of the mask elements to have value $+1$ and set the other half to $-1$, so that $D \hH_t = 0$ by construction. The DSM patterns $h_t$ suitable for use in the CAKE sensing strategy \eqref{eq:cake} are linear combinations of the form
\begin{equation}\label{eq:ht}
h_t = \alpha \hL_k + \beta \hH_t,
\end{equation}
where $k = \floor{(t-1)/B} + 1$, and $t = 1,\ldots,N$. The scalars $\alpha$ and $\beta$ allow additional flexibility in the generation and are normalized such that $\alpha^2 + \beta^2 = 1$. Individually $\hH_t$ and $\hL_k$ are normalized such that $\twonorm{\hH_t}^2 = \twonorm{\hL_t}^2 = n/m$, so this scaling ensures that $\twonorm{h_t}^2 = n/m$ (as $(\hH_t)^\T \hL_t = 0$ by design).  See Fig.~\ref{fig:DSMCCA} for an example mask pattern. The selection of $\alpha$ and $\beta$ essentially trade-off the quality of the intermediate low-resolution estimate, and the final high-resolution reconstruction. Our experiments reveal that a larger weighting on the high-resolution mask results in higher fidelity reconstructions.

The advantage of the DSM design is that we can obtain a coarse estimate of $\fstar$ with negligible computational cost. The remainder of this section details how this is accomplished. Our target is to recover an estimate of the low-rate low-resolution frame sequence
\begin{equation*}
\fL_k := \frac{1}{dB} \sum_{t \in T_k} D \fstar_t.
\end{equation*}
For each low-resolution frame, an estimate $\h{\fL_k}$ of $\fL_k$ can be found easily from a size $m$ convolution with $y_k$. If we denote $\Sigma_k = \FF^{-1} \diag(\sigma_k) \FF$, this estimate is computed via
\begin{equation}
\label{eq:lrestimate}
\h{\fL_k} = \frac{1}{\alpha B} \Sigma_k^\T y_k.
\end{equation} 
We derive \eqref{eq:lrestimate} in Appendix~\ref{app:dsm}, and we detail how this estimate is used to inform an optical flow-based reconstruction algorithm in Section~\ref{sec:algo}.

% ==================================================
% = Reconstruction                                 =
% ==================================================
\section{Video Reconstruction}
\label{sec:algo}

To solve the CS minimization problems \eqref{eq:constrained} or \eqref{eq:l1regularized}, we use well-established gradient-based optimization methods. These iterative algorithms are attractive because they utilize repeated applications of the operators $A$ and $A^\T$ \cite{sparsa,gpsr,SPGL1}. For most CS matrices, these matrix-vector products are a large computational burden and storing such matrices is memory intensive even for modest image sizes. However, the BCCB structure of $A$ allows for computationally fast matrix-vector products using FFTs that do not require explicit storage of the matrix. In our case, the memory footprint is no larger than that of the image itself. In the remainder of this section, we describe how these algorithmic approaches are adapted for video reconstruction as well as describe how optical flow constraints can be introduced to improve reconstruction performance.

\subsection{Reconstruction for Video}

Given measurements from the proposed CAKE imaging system, where we have designed the mask patterns for sparsity in a temporal transform, we recover the frames by solving a minimization similar to \eqref{eq:l1regularized}.
%\begin{equation*}
%\begin{aligned}
%\widehat{\theta} &= \argmin_\theta \tfrac{1}{2} \twonorm{A \theta - y}^2+ \tau \|\theta\|_1 \\
%\widehat{f} &= (W^{-1} \otimes I_n) \widehat{\theta}.
%\end{aligned}
%\end{equation*}
However, since successive video frames are generally highly correlated, there are gains in considering the differences between subsequent frames instead. We therefore define the difference frame sequence $\theta = [\theta_1^\T, \ldots, \theta_N^\T]^\T$ such that
\begin{equation}
    \label{eq:diffframes}
    \theta_t = 
    \begin{cases}
        f_t & t = 1, \\
        f_t - f_{t-1} & t = 2, \ldots, N.
    \end{cases}
\end{equation}
From $\theta$, the frames $f_t$ can easily be recovered by $f_t = \sum_{i=1}^t \theta_i$, $t = 1,\ldots,N$, which can be expressed compactly as $f = (L \kron I_n) \theta$, where $L$ is an $N \times N$ lower triangular matrix of all ones. Additionally, total variation (TV) regularization \cite{ROF} has been shown to lead to increased reconstruction accuracy. We combine these approaches by utilizing a TV penalty on the first frame ($f_1 = \theta_1$), and an sparsity penalty on the subsequent difference fames and solve the following minimization problem
\begin{equation}
    \label{eq:CAKE1}
    \begin{aligned}
        \h{\theta} & = \argmin_{\theta \in \reals^{nN}} &
            & \tfrac{1}{2} \twonorm{A (L \kron I_n) \theta - y}^2 \\
            & & &+ \tau_\text{TV}\|\theta_1\|_\text{TV} 
            + \tau_1 \sum_{t=2}^N \|\theta_t\|_1 \\
        \h{f} &= (L \kron I_n) \h{\theta}.
    \end{aligned}
\end{equation}
Note that in the above we are solving for the entire video sequence in one large optimization problem. This is the reconstruction technique we use in the experimental results section.

Empirically, this leads to better reconstruction performance however this may pose computational challenges for video sequences of longer duration. In such cases, an effective approach is to process the video in an online fashion, solving for only a few blocks of frames simultaneously. In many applications, such as surveillance and monitoring, the video frames are strongly correlated, and the solution to a previous block of frames may be used as an initialization to the algorithm to solve the next block of frames. This estimate will generally be very accurate, and therefore, relatively few iterations are needed to obtain a solution to each optimization problem. In previous work, we noticed that when the amount of processing time allotted per frame is held constant, the accuracy generally increases with the number of frames processed simultaneously.  However, one simply cannot solve for arbitrarily many frames simultaneously, as the improvement in accuracy diminishes when the size of the problem is such that only a very small number of reconstruction iterations can be run within the allotted time.  We refer the reader to \cite{MarciaEUSIPCO} for details.

\subsection{Reconstruction using Estimated Optical Flow}
\label{sec:opticalflow}

Optical flow is a measurement of the spatial motion of the grayscale intensity of a video sequence over time \cite{verri1989} and has enjoyed success in the computer vision community for several applications such as video coding \cite{moulin1997multiscale-modeling-and-estimation-of-motion}, and robot vision \cite{song2011a-kalman-filter-integrated-optical-flow-method}. The development of fast algorithms for estimating optical flow, such as \cite{liu2009phd} or \cite{ayvaci2010} based on convex optimization,  has enabled its use in the signal processing community. In addition to its use in this work, we have successfully employed optical flow in related work to design adaptive sensing matrices based on compressive coded apertures \cite{harmany2011motion-adaptive}.

Mathematically, optical flow is a time-varying vector field $v_t$, $t=1,\ldots,N$ over the image space which describes how the grayscale intensity of frame $f_t$ is propagated to frame $f_{t+1}$. For two-dimensional video sequences, we consider $v_t = (v_{1,t}, v_{2,t})$, where $v_{1,t} \in \reals^n$ is the horizontal component and $v_{2,t} \in \reals^n$ is the vertical component of the optical flow. We can express the effects of the optical flow $v_t$ as a matrix $V_t$ such that $f_{t+1} \approx V_t f_{t}$. Grouping these constraints over the duration of the video sequence, we can express this as $Vf \approx 0$, where 
\begin{equation}
    \label{eq:Mmatrix}
    V = \begin{bmatrix}
            V_1     & -I        &           &           & \\
                    & V_2       & -I        &           & \\
                    &           & \ddots    & \ddots    & \\
                    &           &           & V_{N-1}   & -I 
        \end{bmatrix}.
\end{equation}

In this work, we use optical flow as a method to enforce smooth motion fields over an estimated video sequence. Our approach is based on the CS-MUVI approach \cite{cs-muvi}. The reconstruction procedure follows the following steps.
\begin{enumerate}
\item From the low-rate coded data $y$, we compute the low-resolution estimate in \eqref{eq:lrestimate}.
\item We upsample $\h{\fL_k}$, $k=1,\ldots,M$ to match the frame rate and resolution of the sequence $f_t$, $t=1,\ldots,N$. In our implementation, we use spline interpolation based on the \texttt{interp3} MATLAB command.
\item The optical flow field is computed using software provided by Liu \cite{liuopticalflow}, and described in detail in \cite{liu2009phd}. 
\item Using this estimated optical flow, we form the matrix $V$ of motion maps in \eqref{eq:Mmatrix}. Using these motion maps in conjunction with the data, we solve for a set of sparse frame coefficients defined in \eqref{eq:diffframes} via the convex program
%\begin{equation}\label{eq:CAKEOptFlow}
%	\ConMinProb{f \in \reals^{nN}, \theta \in \reals^{nN}}{\onenorm{\theta}}
%	{ f = W \theta, \\
%		& & &\twonorm{AW \theta - y} \le \epsilon_1, \\
%	  & & &\twonorm{ M_t f_t - f_{t+1} } \le \epsilon_2,\\
%	  & & &\quad t = 1, \ldots, N-1} \\
%\end{equation}
\begin{eqnarray}\label{eq:CAKEOptFlow}
	\underset{\theta \in \reals^{nN}}{\textrm{minimize}} \hspace{.1cm} && \hspace{-.2cm} {\onenorm{W^T\theta}} \\
	\textrm{subject to\ \ } \hspace{-.2cm} 
      &&  \hspace{-.2cm} \twonorm{A (L \otimes I_n) \theta - y} \le \epsilon_1,  \nonumber \\
	  && \hspace{-.2cm} \twonorm{ V(L \otimes I_n) \theta } \le \epsilon_2, \ \  t = 1, \ldots, N\!-\!1 \nonumber
\end{eqnarray}
where $W$ is a sparsity basis for the difference frame sequence $f$.
\end{enumerate}

In our implementation, we consider sparsity in the Daubechies-4 wavelet basis. Here $\epsilon_1$ and $\epsilon_2$ are tuning parameters that dictate how tightly the data and optical flow constraints should be enforced. The convex program in above can be placed into a standard $\ell_2$-constrained $\ell_1$ program which can be solved using \texttt{SPGL1} \cite{SPGL1}. The added optical flow constraints are an effective tool for enforcing smooth motion in the video sequence as well as constraining the optimization program which, in our experiments, results in faster convergence behavior. We demonstrate the effectiveness of this method in Section~\ref{sec:experiments}.

% ==================================================
% = Numerical Experiments                          =
% ==================================================
\section{Numerical Experiments}
\label{sec:experiments}

\begin{figure}[t]
\begin{center}
\includegraphics[width=\columnwidth]{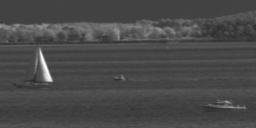}
\caption{Example frame (frame 21) from the ground truth video sequence used in the numerical experiments.}
\label{fig:singleframe} 
\end{center}
\end{figure} 

\newlength{\reconfigsize}
\setlength{\reconfigsize}{2.1cm}

\begin{figure*}[htbp]
\centering \small
\begin{tabular}{m{1.8cm} m{\reconfigsize} m{\reconfigsize} m{\reconfigsize} m{\reconfigsize} m{\reconfigsize}}
	& 	  	&  \centering Conventional  	& & \centering Dual-Scale	& \hspace{.65cm} Optical 
\\
  		& \centering Truth & \centering Spline & \centering CAKE &  \centering Mask  
	& \hspace{.8cm}  Flow\\
	& 		&  \centering Upsampling	&		& \centering CAKE		
	& \hspace{.7cm} CAKE  \\
	Frame 5 & 
	\includegraphics[width=\reconfigsize]{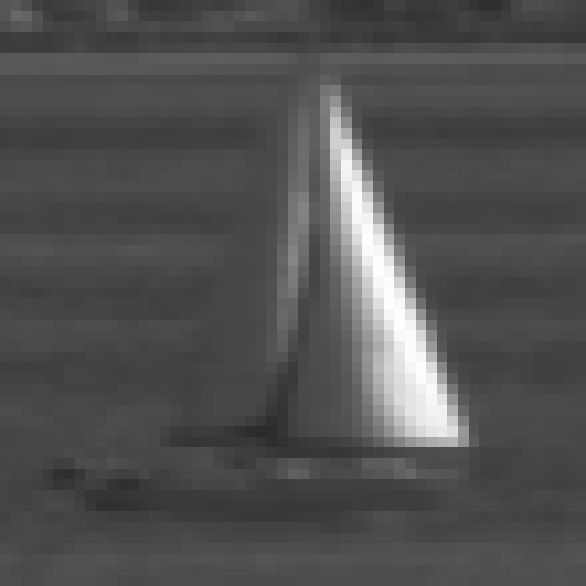}  &
	\includegraphics[width=\reconfigsize]{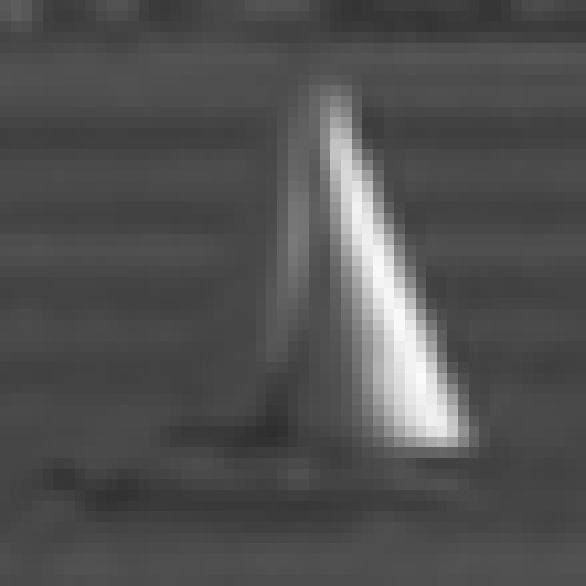}  & 
	\includegraphics[width=\reconfigsize]{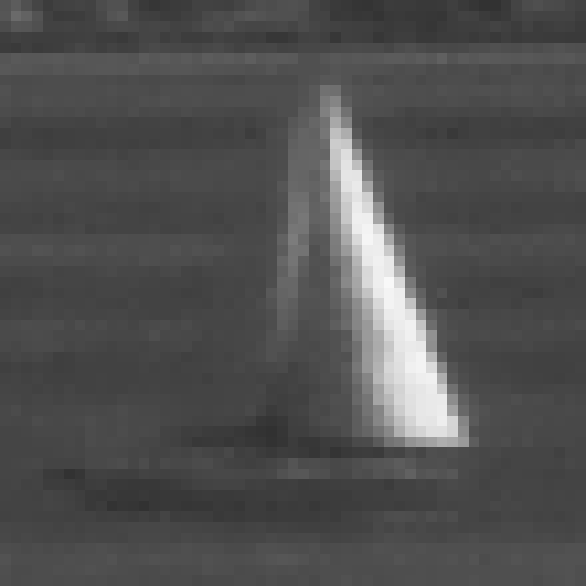} &
	\includegraphics[width=\reconfigsize]{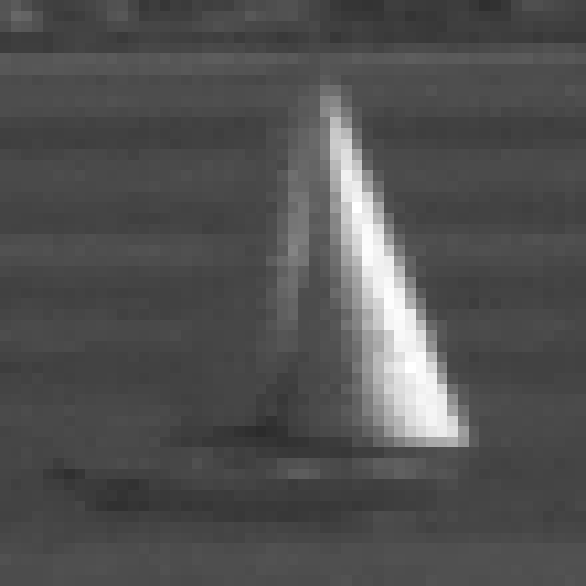} & 
	\includegraphics[width=\reconfigsize]{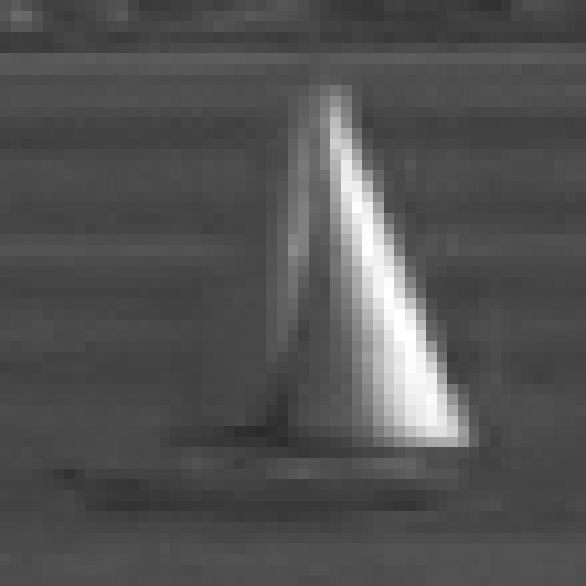}  
	\\
	\\[-.3cm]
	Residual
	&
	&
	\includegraphics[width=\reconfigsize]{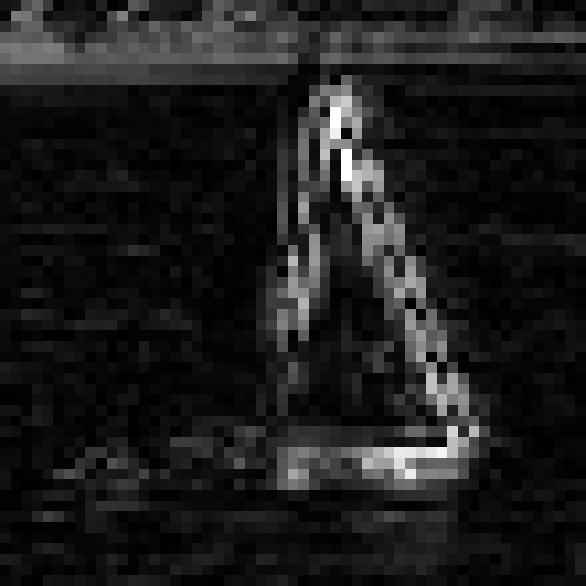}  & 
	\includegraphics[width=\reconfigsize]{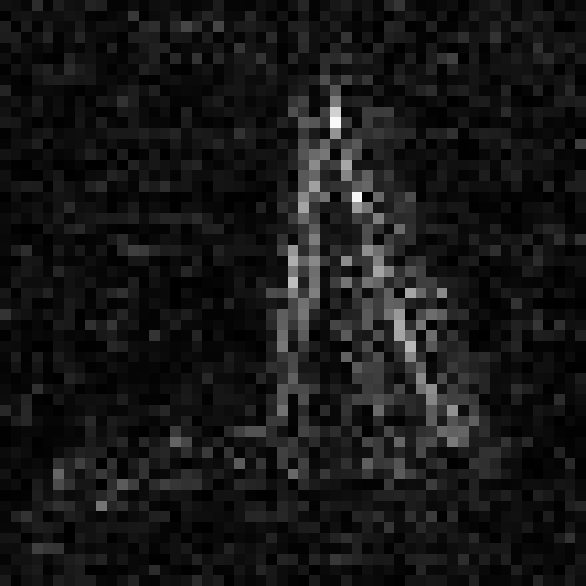} &
	\includegraphics[width=\reconfigsize]{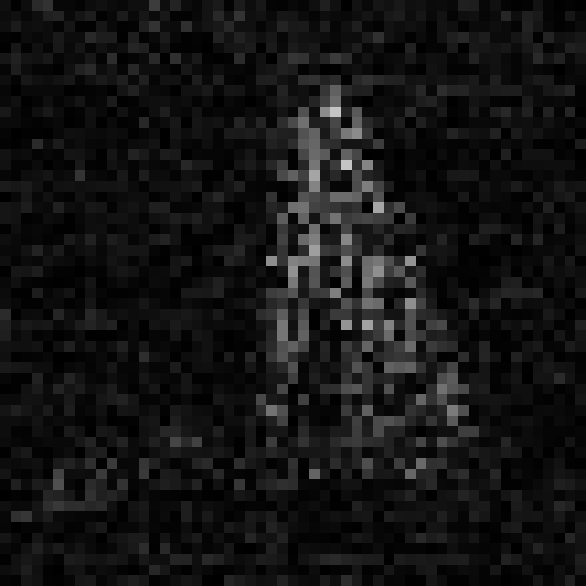} & 
	\includegraphics[width=\reconfigsize]{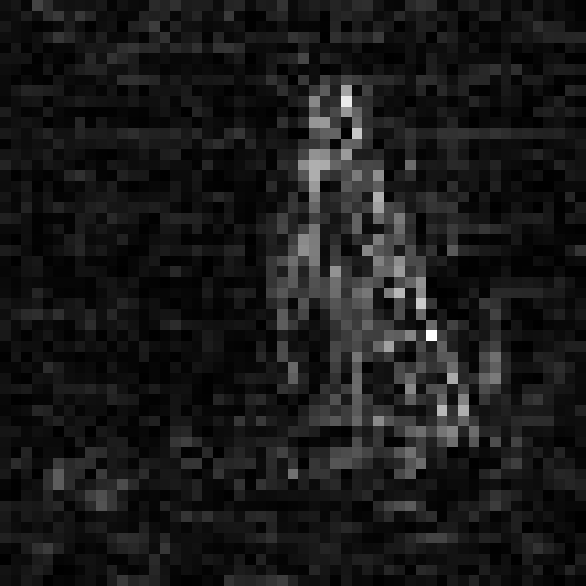}  
	\\
	\\[-.3cm]
	\begin{tabular}{ll}
	\hspace{-.4cm} Difference \\
	\hspace{-.4cm} between \\
	\hspace{-.4cm} Frame 5 and \\
	\hspace{-.4cm} Frame 6
	\end{tabular}
	&
	\includegraphics[width=\reconfigsize]{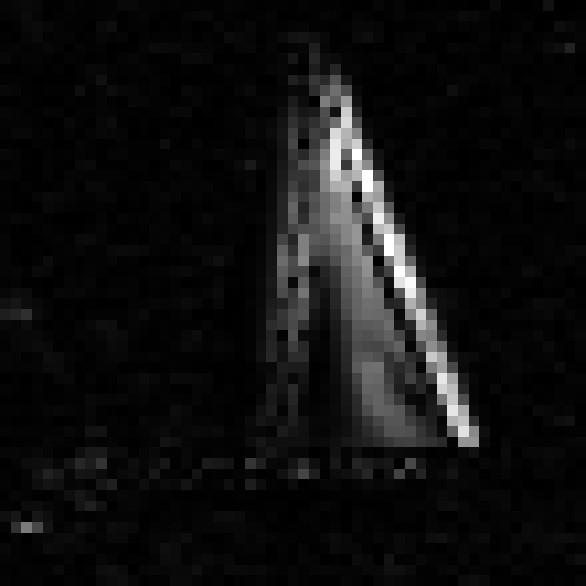}  & 
	\includegraphics[width=\reconfigsize]{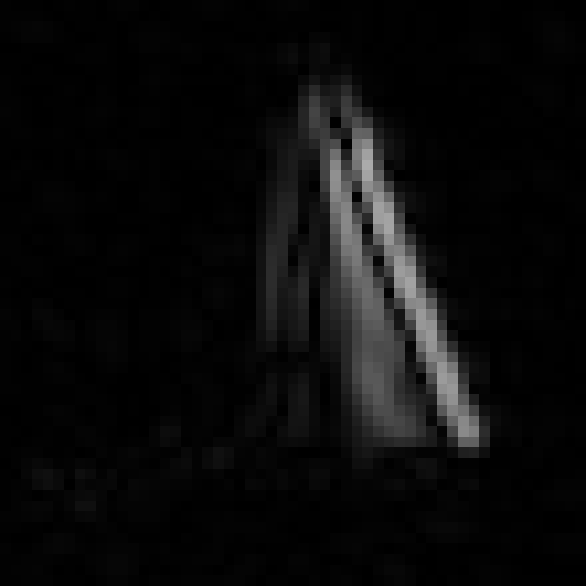}  & 
	\includegraphics[width=\reconfigsize]{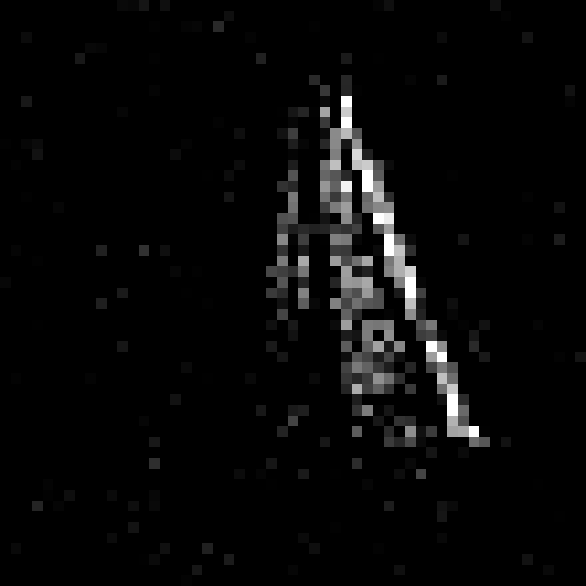} &
	\includegraphics[width=\reconfigsize]{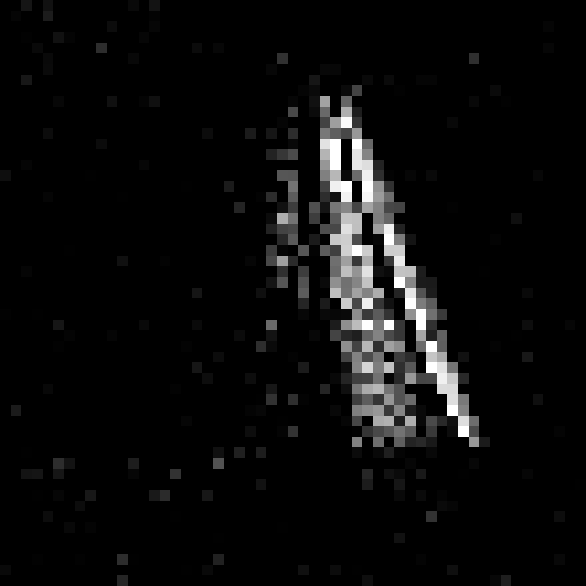} & 
	\includegraphics[width=\reconfigsize]{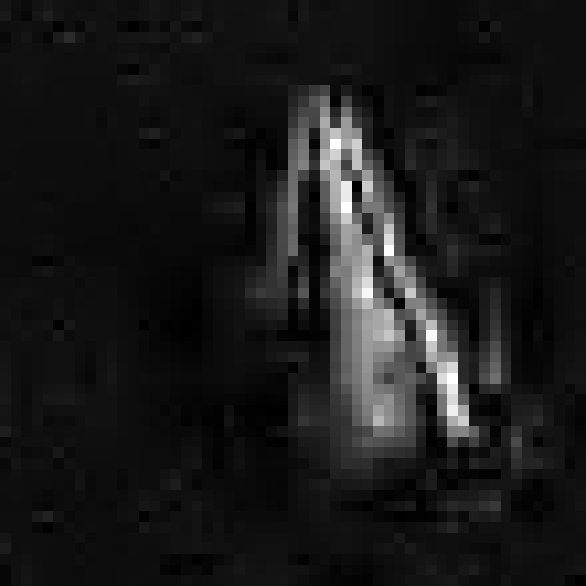}  
	\\
	\\[-.3cm]
	Frame 24 &
	\includegraphics[width=\reconfigsize]{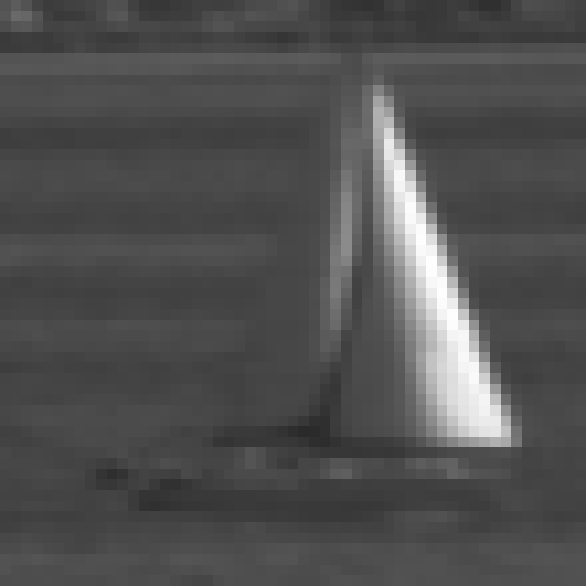}  &
	\includegraphics[width=\reconfigsize]{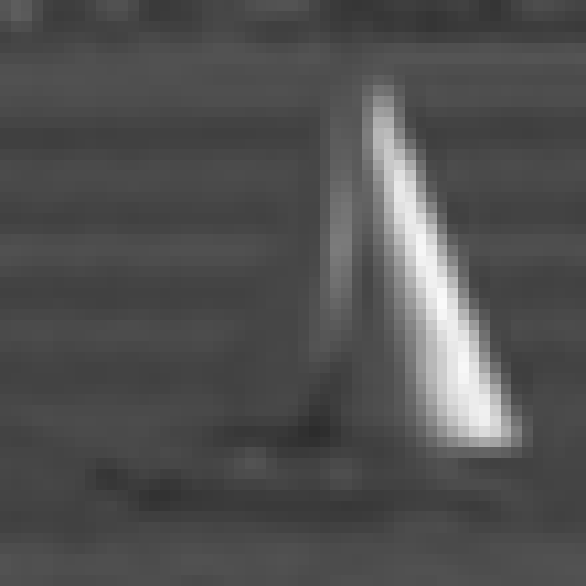}  & 
	\includegraphics[width=\reconfigsize]{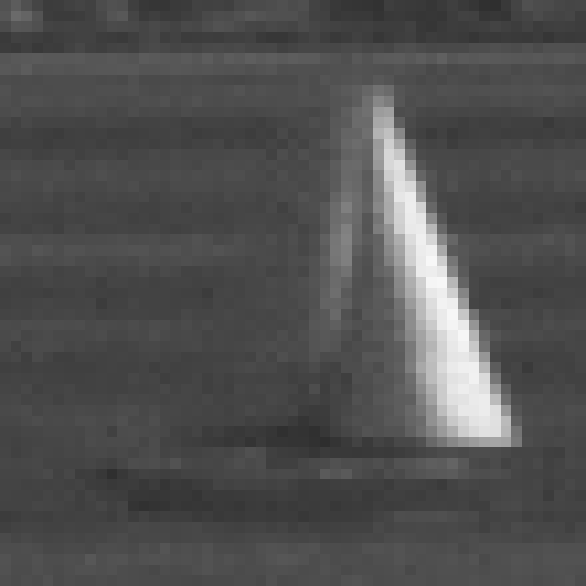} &
	\includegraphics[width=\reconfigsize]{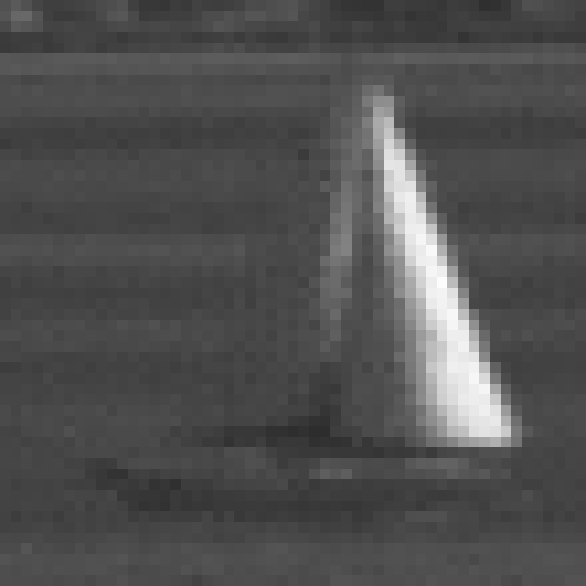} & 
	\includegraphics[width=\reconfigsize]{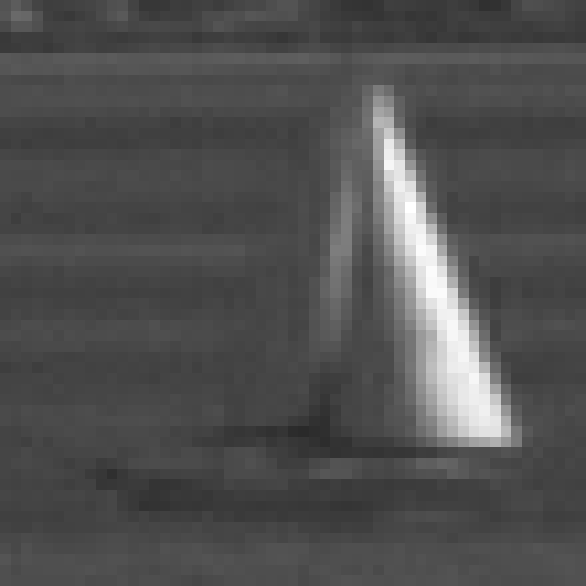}  
	\\
	\\[-.3cm]
	Residual &
	&
	\includegraphics[width=\reconfigsize]{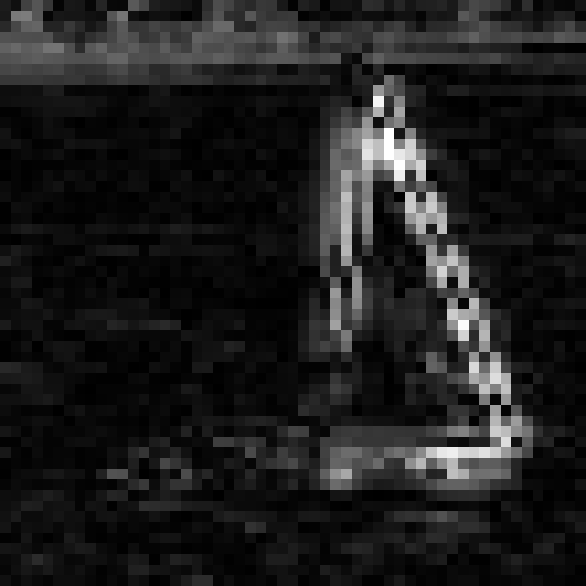}  & 
	\includegraphics[width=\reconfigsize]{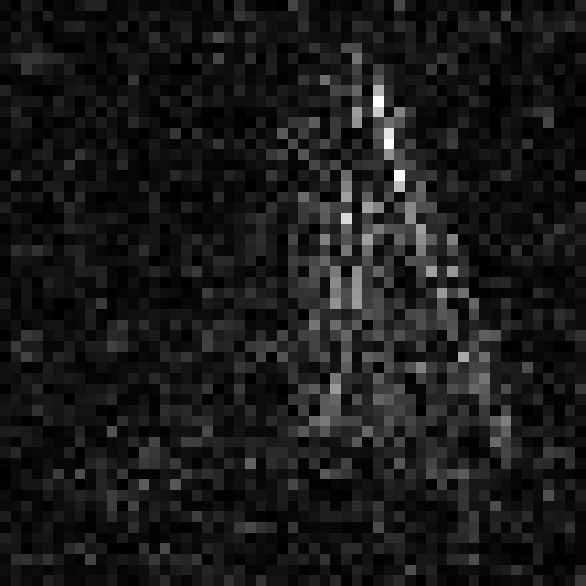} &
	\includegraphics[width=\reconfigsize]{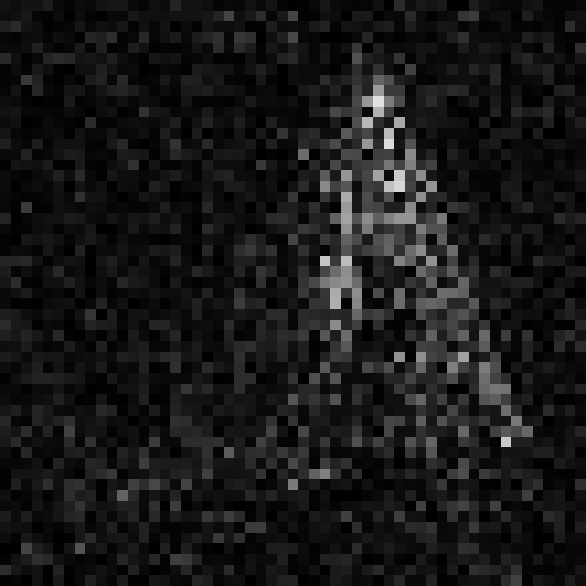} & 
	\includegraphics[width=\reconfigsize]{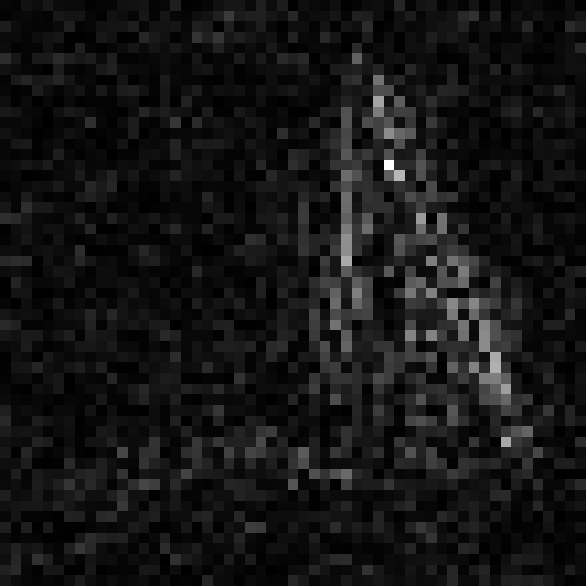}  
	\\
	\\[-.3cm]
	\begin{tabular}{ll}
	\hspace{-.4cm} Difference \\
	\hspace{-.4cm} between \\
	\hspace{-.4cm} Frame 24 and \\
	\hspace{-.4cm} Frame 25
	\end{tabular}
	&
	\includegraphics[width=\reconfigsize]{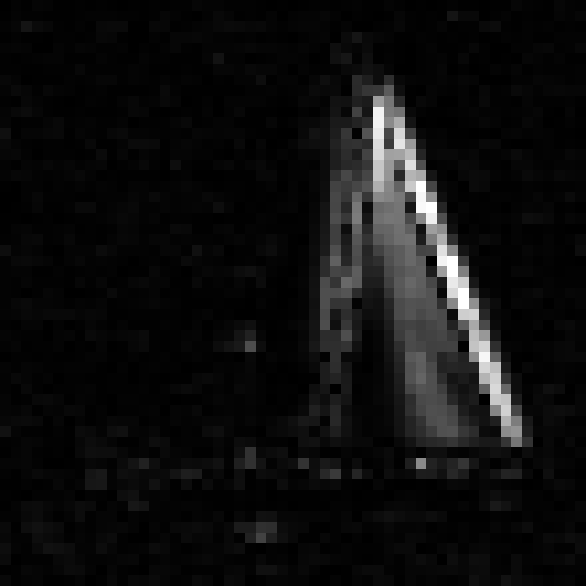}  & 
	\includegraphics[width=\reconfigsize]{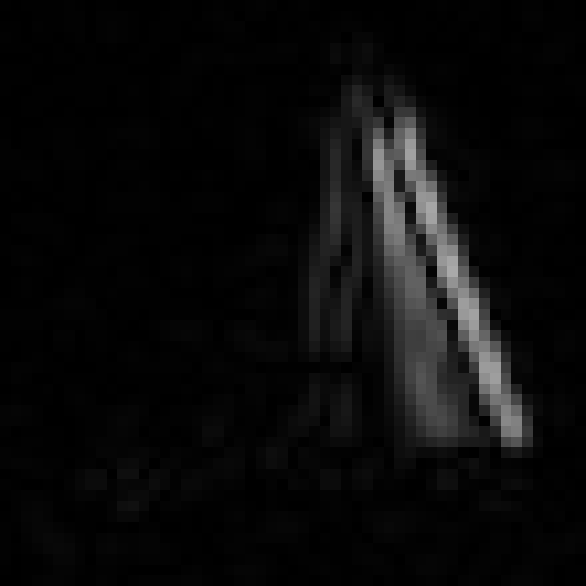}  & 
	\includegraphics[width=\reconfigsize]{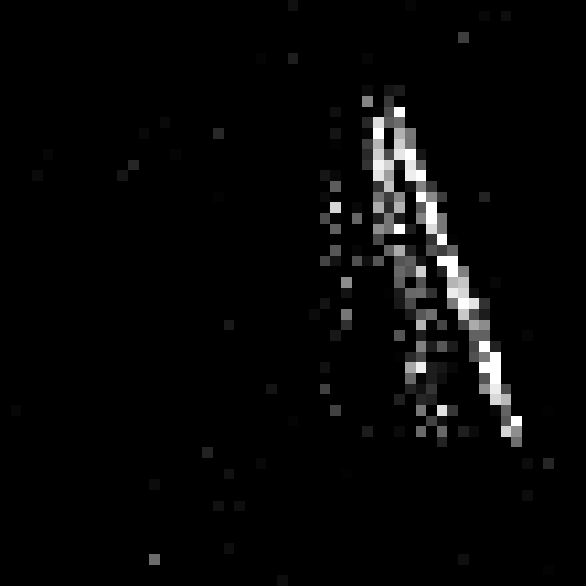} &
	\includegraphics[width=\reconfigsize]{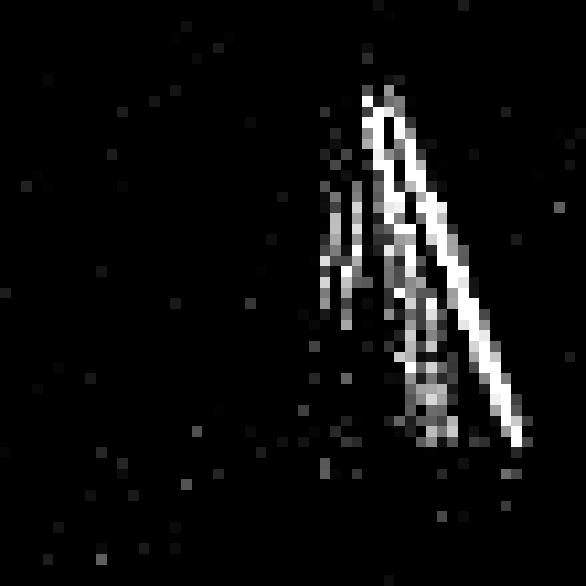} & 
	\includegraphics[width=\reconfigsize]{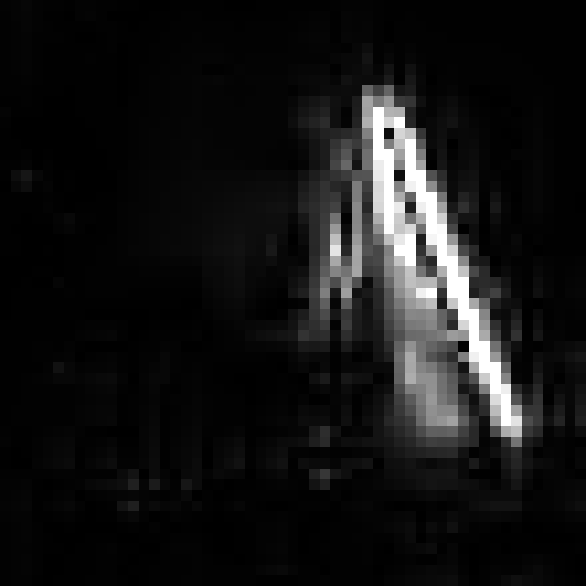}  
	\\
	\\[-.3cm]
	Total RMSE & &
	\centering \ 0.7192\% & 
	\centering \ 0.5270\% & 
	\centering \ 0.4764\% & 
	\centering \ 0.4503\%
\end{tabular}
\caption{Results obtained for the sailboat region of interest for an example pair of frames (Frames 5 and 24). Shown are the true frames and the reconstructions from traditionally sampled data, from CAKE sensed data, from CAKE using a dual-scale mask, and CAKE with optical flows. Also reported are the total RMSEs from Frames 5 to 24 for each approach --- the average over 10 trials are listed in Table I.  For comparison we show the magnitude of the residuals as compared with the truth as well as the difference between consecutive frames.   Note the overall better scene reconstruction both spatially and temporally using the three CAKE methods.  Specifically, they more accurately recover the edges of the main sail.  Furthermore, the low amplitude and blurry frame differences using spline upsampling indicate that spline upsampling captures the movement of the boat less accurately than CAKE.  Finally, using CAKE with optical flow results in much less noisy inter-frame differences, which shows that this method reconstructs motion to a higher accuracy.}
%\todo{note that using the splines, the temporal detail is lost, and hence we are getting lower temporal resolution. Note that using the optical flow, the inter-frame differences is much less noisy, which shows that this method reconstructs motion to a higher accuracy. (may shift to text vs caption).}}
\label{fig:SailboatPanel} 
\end{figure*}

In this section, we demonstrate the effectiveness of the proposed CAKE architecture at successfully recovering a video sequence of short-wave infrared (SWIR) data collected (courtesy of Jon Nichols at NRL) by a short-wave IR ($0.9-1.7\mu\text{m}$) camera. The camera is based on a $1024\times 1280$ InGaAs (Indium, gallium
arsenide) focal plane array with $20\mu\text{m}$ pixel pitch.  Optically, the camera was built around a fixed, $f/2$ aperture and provides a $6^{\circ}$ field of view along the diagonal with a focus range of $50\text{m}\rightarrow\infty$.  Imagery were output at the standard 30Hz frame rate with a 14 bit dynamic range. An example frame is shown in Fig.~\ref{fig:singleframe}. In this sequence, the three boats are traveling at different velocities with respect to the slowly-changing background of the waves. The size of each frame is $128 \times 256$, and we consider reconstructing 28 frames of the sequence.

We consider CAKE observations where we downsample spatially by a factor of 2 in both directions ($d_1 = d_2 = 2$), and use a coded exposure block length of $B = 4$. We compare three CAKE paradigms: single mask, dual-scale masks, and CAKE using optical flow.  We reconstruct the entire video sequence using all $M = N/B = 7$ low-resolution low-rate frames using a total variation penalty on the first frame, and $\ell_1$ sparsity penalty on all subsequent difference frames. We optimize the regularization parameters to minimize the reconstruction error. Specifically, for the single- and dual-scale mask CAKE  \eqref{eq:CAKE1}, $\tau_{\textrm{TV}} \! = \! 1.0 \! \times \! 10^{-2}$, $\tau_1 \! = \! 2.0 \! \times \! 10^{-2}$, and for CAKE with optical flow \eqref{eq:CAKEOptFlow}, $\epsilon_1 \! = \!  4.3\times 10^{-2}$ and $\epsilon_2 = 4.3\times 10^{3}$.
We used the weights $\alpha = 0.383$ and $\beta = 0.924$ in \eqref{eq:ht}.  For comparison, we consider traditionally captured data (\ie by simply averaging over $d_1 \times d_2 \times B$ blocks of the spatiotemporal video sequence). To interpolate this data to the original resolution of the video sequence, we consider spline interpolation.

We show the estimates for the different acquisition and reconstruction methods in Fig.~\ref{fig:SailboatPanel}. Here we focus only on a region of interest (ROI) of the sailboat. We see that the CAKE sensing is able to reconstruct the scene with a higher spatial and temporal resolution. This is evidenced in the residuals, which include much less critical scene structure than the conventional system. We see from the examination of the difference frame that the nearest-neighbor reconstruction from conventionally sampled data yields only slight motion over the two frames and suffers from poor spatial resolution. Using spline interpolation helps improve the spatial resolution, but it is still insufficient to recover the scene with high temporal resolution, as can be seen in the blur on the leading edge of the sail. Numerically we quantify the performance over the entire video sequence in terms of the RMSE (\%), $100 \cdot \|\widehat{f} - \fstar\|_2/\|\fstar\|_2$, calculated  over the sailboat ROI in order to effectively assess ability to capture motion information in the scene. This is tabulated in Table~\ref{tab:CAKERMSE}.  In summary we see that the CAKE acquisitions are able to outperform traditionally sampled video in terms of reconstruction accuracy and reconstructing salient motion.
On average, to reconstruct the 28 video frames, CAKE took 96 minutes and Dual-Scale Mask CAKE took 93 minutes, while CAKE with Optical Flow took only 31 minutes.  Conventional spline upsampling took nearly 23 seconds.  

It should be noted that there are limitations to the CAKE system in the presence of strong motion. In this case, the sparsity level of the difference frames may drastically increase as the previous frame ceases to be a good prediction of the next frame. As such, the RIP bound in Theorem~\ref{thm:cake} states the number of measurements we require to reconstruct the scene must necessarily increase, requiring either a faster temporal resolution measurements or higher resolution FPA to achieve the same accuracy. Because of this balance, a system designer may need to make important engineering tradeoffs to implement CAKE acquisition for a particular application.

\begin{table}[t] \label{tab:CAKERMSE}
\centering
\begin{tabular}{lr}
\toprule
Sensing Architecture & Reconstruction RMSE \\
& (Average over 10 Trials) \\
\hline
% Conventional Spline Upsampling 1-trial: 0.7192\%
Conventional Spline Upsampling & 0.7192\% \\
% CAKE 1-trial: 0.5268\%
CAKE & 0.5270\% \\
% Dual-Scale Mask CAKE 1-trial: 0.4736\%
Dual-Scale Mask CAKE & 0.4764\% \\
% Optical Flow CAKE 1-trial: 0.4478\%
Optical Flow CAKE & 0.4503\% \\
\hline
\end{tabular}
\caption{Reconstruction RMSE achieved for the conventional and coded aperture and keyed exposure (CAKE) architectures over the video sequence. Results are reported for the region of interest of the sailboat. Due to boundary issues, we report the RMSE discounting the first and last block of 4 frames. %\todo{add column for RMSE for particular expeirment} 
}
\end{table}

% ==================================================
% = Hardware and Practical Implementation          =
% = Considerations                                 =
% ==================================================
\section{Hardware and Practical Implementation Considerations}
\label{sec:hardware}

In this section we describe how we shift from modeling the coded aperture masks in a way that is compatible with compressed sensing theory, to a model that describes their actual implementation in an optical system. In particular, gaps between theory and practice arise due to:
\begin{itemize}
\item when using lenses with bandwidth below the desired resolution;
\item system calibration issues;
\item nonnegativity of scenes and sensing matrices in incoherent light settings;
\item inaccurate synchronization of spatial light modulators and sensor read-out;
\item low photon efficiency of some architectures;
\item hardware implementation of downsampling operation;
\item quantization errors in observations.
\end{itemize}
Each of these is described in more detail below.

{\bf Lens bandwidth:} Our analysis does not account for the bandwidth of the lenses; in particular, we implicitly assume that the bandwidth of the lenses is high enough that band-limitation effects are negligible at the resolution of interest.

{\bf System calibration issues:}  In all of the hardware settings described below, precise alignment of the optical components (\eg the mask and the focal plane array) is critical to the performance of the proposed system. Often a high-resolution FPA is helpful for alignment and calibration. Even when we have the ability to estimate $A$ precisely,
there are settings where using an approximation of $A$ has advantages;
for instance, when we can approximately compute $Af$ using fast
Fourier transforms, then conducting sparse recovery is much faster
than with a dense matrix representation of $A$.
When we run a sparse recovery algorithm with  an inaccurate sensing matrix $A$, it corresponds to the  observation model
$$y = Af + Ef + w,$$ where $Ef$ represents the difference between the
{\em true} projections collected by  imager and the {\em
assumed} projections in $A$. The term $E f$ can be thought of as
signal-dependent noise. Recent work analyzes  the theoretical
ramifications of these kinds of errors \cite{LohWai12}.

{\bf Nonnegativity in incoherent light settings:} In this paper we focus on {\em incoherent} light settings (consistent with many applications in astronomy, microscopy, and infrared imaging). In this case, the coded aperture must be real-valued and flux-preserving (\ie the light intensity hitting the detector cannot exceed the light intensity of the source).  In a conventional lensless coded aperture imaging setup, the point spread function associated with the aperture is the mask pattern $h$ itself. 
% For a binary mask pattern, this amounts to restricting $h \in \{0, 1/n\}^n$.
To shift a RIP-satisfying aperture as in Theorem \ref{thm:cake} to an implementable aperture, one simply needs to apply an affine transform to $h$ mapping $[-\sqrt{d/n},\sqrt{d/n}]$ to $[0, 1/n]$. This transform ensures that the resulting mask pattern is nonnegative and flux-preserving. In previous work \cite{Marcia:09,MarciaHW_SPIE2010}, we discuss accounting for nonnegativity during reconstruction by a mean subtraction step.

{\bf Synchronization issues:} Programmable spatial light modulators (SLMs) allow changing the mask pattern over time, a critical component of the CAKE concept. However, in order for the proposed approach to work, the mask or SLM used must be higher resolution than the FPA. Currently, very high resolution SLMs are still in development. Chrome on quartz masks can be made with higher resolution than many SLMs, but cannot be changed on the fly unless we mount a small number of fixed masks on a rotating wheel or translating stage. Furthermore, the changing of the mask patterns must be carefully synchronized to th timing of readout on the focal plane array.

{\bf Photon efficiency:} Phase-shift masks for coded aperture imaging have been implemented recently using a phase screen \cite{phaseCode}. This approach allows one to account for diffraction in the optical design. However, depending on the precise optical architecture, phase shift masks may be  less photon-efficient than amplitude modulation masks. 

{\bf Downsampling in hardware:} In developing RIP-satisfying coded apertures, Theorem \ref{thm:cake} assumes the subsampling operation $S$ selects one measurement per $d_1 \times d_2$ block. From an implementation standpoint, this operation effectively discards a large portion ($(d-1)/d$) of the available light, which would result in much lower signal-to-noise ratios at the detector. A more pragmatic approach is to use larger detector elements that essentially sum the intensity over each $d_1 \times d_2$ block, making a better use of the available light. We call this operation \emph{integration downsampling} and denote it by $D$ to distinguish it from subsampling. The drawback to this approach is that we lose many of the desirable features of the system in terms of the RIP. Integration downsampling causes a large coherence between neighboring columns of the resulting sensing matrix $A$. An intermediate approach would randomly sum a fraction of the elements in each size $d$ block, which increases the signal-to-noise ratio versus subsampling, but yields smaller expected coherence. This approach is motivated by the random demodulation proposed in \cite{RombergToeplitz} and described in Sec~\ref{sec:dsm}, whereby the signal is multiplied by a random sequence of signs $\{-1, +1\}$, then block-wise averaged. The pseudo-random summation proposed here can be thought of as an optically realizable instantiation of the same idea where we multiply by a random binary $\{0, 1\}$ sequence.  

{\bf Noise and quantization errors:} While CS is particularly useful when the FPA needs to be kept compact, it should be noted that CS is more sensitive to measurement errors and noise than more direct imaging techniques.  The experiments conducted in this paper simulated very high signal-to-noise ratio (SNR) settings and showed that CS methods can help resolve high resolution features in images. However, in low SNR settings CS reconstructions can exhibit significant artifacts that may even cause more distortion than the low-resolution effects associated with conventional coded aperture techniques such as MURA.
Similar observations are made in \cite{CSvsConventional}, which presents a direct comparison of the noise robustness of CS in contrast to conventional imaging techniques both in terms of bounds on how reconstruction error decays with the number of measurements and in a simulation setup; the authors conclude that for most real-world images, CS yields the biggest gains in high signal-to-noise ratio (SNR) settings. Related theoretical work in \cite{raginskyWillettPCS_TSP} show that in the presence of low SNR photon noise, theoretical error bounds can be large, and thus the expected performance of CS may be limited unless the number of available photons to sense is sufficiently high. These considerations play an important role in choosing the type of downsampling to implement.
Similar issues arise when considering the bit-depth of focal plane arrays, which corresponds to measurement quantization errors. Future efforts in designing optical CS systems must carefully consider the amount of noise anticipated in the measurements to find the optimal tradeoff between the focal plane array size and image quality.

%Both Robucci {\it et al.} \cite{Robucci:10} and Majidzadeh {\it et al.} \cite{Jacques:10} have proposed performing the analog, random convolution step in complementary, metal-oxide-semiconductor (CMOS) electronics.  A clear advantage to this architecture is that the additional optics required for spatial light modulation are removed in favor of additional circuitry, immediately reducing imager size.  

%Additionally, the mask generation distribution described in Eq.~\eqref{eq:psmdist} will result in negative entries for the corresponding PSF $\hup = \FF^{-1}\sigma$. To compensate for this, the phase mask must be mean-shifted to make all entries nonnegative, so that we are actually implementing
%\begin{equation}
%\hup_+ = c(\hup - \min(\hup)\ones),
%\end{equation}
%with the constant $c$ selected so that the implementable PSF $\hup_+$ is flux-preserving and $\ones$
%is a matrix of ones and of the same dimension as $\hup$. 

% ==================================================
% = Conclusions                                    =
% ==================================================
\section{Conclusions}
\label{sec:conc}

%
%Compressed sensing offers a strong theoretical foundation for sparse signal recovery. However, practical and implementable imaging system designs based on CS theory have lagged far behind.  
In this paper, we demonstrate how CS principles can be
applied to physically realizable optical system designs, namely coded aperture imaging.
Numerical experiments show that CS methods can help resolve high resolution
features in videos that conventional imaging systems cannot.
%Also, while the Fourier-domain phase-shift mask \cite{RombergToeplitz} slightly outperforms our 
%proposed compressive coded aperture amplitude modulation mask,
%the phase-shift mask model requires a very high resolution spatial light modulator and is
%less photon efficient.  
%Future efforts will explore mask designs with this consideration in mind.
We have also demonstrated that our CAKE acquisition system can recover video sequences from highly under-sampled data in much broader settings than those considered in initial coded exposure studies. However, our derived theoretical limits show that there are important tradeoffs involved that depend on the spatial sparsity and temporal correlations in the scene that ultimately govern the accuracy of our reconstructions for a specified number of compressive measurements.
Finally, we note that our proposed approach performs well 
in high signal-to-noise ratio settings, but like all CS reconstructions, 
it can exhibit significant artifacts in high-noise settings.

%Known CS matrices are often zero-mean, which
%means that some matrix elements are negative.  In an imaging context,
%this is equivalent to light intensity \textsl{subtraction} within the
%projection operator, which is physically unrealistic and can result in negative-valued 
%observations.  

Two practical issues associated with coded aperture imaging in general are the blur due to misalignment of the mask and diffraction and interference effects. Non-compressive aperture codes have been developed to be robust to these effects \cite{diffractionEffects}. One important avenue for future research is the development of {\em compressive} coded aperture makes with similar robustness properties. Non-compressive coded apertures have also been shown useful in inferring the depth of different objects in a scene; similar inference may be possible with the compressive coded apertures described in this paper. However, using the current proof techniques here or in \cite{krahmermendelson13suprema} cannot be directly applied as the additional structure in the DSM masks results in random vectors that are not isotropic sub-Gaussian random vectors.

% ==================================================
% = Acknowledgements                               = 
% ==================================================
\section*{Acknowledgements}
The authors would like to thank Prof. Justin Romberg and Dr. Kerkil Choi for several valuable discussions, Dr. Aswin Sankaranarayanan for providing code to reproduce the results in \cite{cs-muvi}, and Dr. Seda Senay for assisting with the numerical experiments.

% ==================================================
% = Appendix                                       =
% ==================================================
\appendix

\subsection{Proof of the RIP for CAKE Sensing}
\label{sec:proof}
Here we present the proof of Theorem~\ref{thm:cake} that our CAKE sensing architecture satisfies the restricted isometry property. The proof uses the same techniques as that of Theorem 4 in \cite{haupttoeplitz}, where the RIP is established by shifting the analysis of the submatrices of $A$ to the entries of the Gram matrix $G = A^\T \! A$ by invoking Ger\v{s}gorin's disc theorem \cite{Verga}. This theorem states that the eigenvalues of an $n \times n$ complex matrix $G$ all lie in the union of $n$ discs $d_j(c_j,r_j)$, $j = 1,2,\ldots,n$, centered at $c_j = G_{j,j}$ with radius
\begin{equation*}
r_j = \sum_{i=1 \atop i \ne j}^m |G_{i,j}|.
\end{equation*}
In essence, we show that with high probability $G \approx I$, so that the eigenvalues of $G$ are clustered around one with suitably high probability. 

\begin{IEEEproof}[Proof of Theorem~\ref{thm:cake}]
Note that the Gram matrix has a certain block structure; since $A = [A_1 \cdots A_B]$, we have
\begin{equation*}
G = A^\T A = 
\begin{bmatrix}
	A_1^\T A_1    & A_1^\T A_2    & \cdots    & A_1^\T A_B  \\
	A_2^\T A_1    & A_2^\T A_2    & \cdots    & A_2^\T A_B  \\
	\vdots       & \vdots       & \ddots    & \vdots     \\
	A_B^\T A_1    & A_B^\T A_2    & \cdots    & A_B^\T A_B
\end{bmatrix}.
\end{equation*}

In the interest of notational simplicity, it is easier to work on the two-dimensional images versus their one-dimensional vectorial representations. This means that for our coded aperture mask patterns $h_t$, instead of our previous indexing notation $[h_t]_k$, we use two dimensional indexing $[h_t]_{(k_1,k_2)}$ such that
\begin{equation*}
[h_t]_{(k_1,k_2)} = [h_t]_{k_1 + n_1 k_2} = [h_t]_k.
\end{equation*}
Additionally, for the sensing matrices $A_t$, we use the indexing
\begin{equation*}
[A_t]_{(l_1,l_2),(k_1,k_2)} = [A_t]_{l_1 + m_1 l_2,k_1 + n_2 k_2} = [A_t]_{l,k},
\end{equation*}
also define $(k)_n = k \mod n$ for modular arithmetic. 

We first focus on the diagonal blocks of this matrix. First we need a relationship between the matrices $A_t$ and the mask patterns $h_t$. It can be shown that
\begin{equation*}
\begin{aligned}
\left[A_t\right]_{l,k} 
= 
[h_t]_{\big(
( (l_1-1)d_1 - k_1 + 1 )_{n_1}+1,
( (l_2-1)d_2 - k_2 + 1 )_{n_2}+1 \big)},
\end{aligned}
\end{equation*}
so the entries of the diagonal blocks of $G$ are
\begin{equation}
\begin{aligned}
\label{eq:grammatrix}
[A_t^\T A_t]_{p,q} \ &\!\! 
= \sum_{l_1 = 1}^{n_1/d_1}\sum_{l_2 = 1}^{n_2/d_2} \\
&h_{ \big( ((l_1-1)d_1 - p_1 + 1)_{n_1}+1,
     ((l_2-1)d_2 - p_2 + 1)_{n_2}+1 \big) }  \\
&\cdot h_{ \big(((l_1-1)d_1 - q_1 + 1)_{n_1}+1,
     ((l_2-1)d_2 - q_2 + 1)_{n_2}+1 \big) }.
\end{aligned}\end{equation}
From the normalization, it is straightforward to show that $\expect [A_t^\T A_t] = I$. Now we need to bound the deviation about the mean via concentration. The diagonal terms are simply a sum of $n = n_1n_2$ bounded \iid entries, and hence for all $t$ and $q$, Hoeffding's inequality yields
\begin{equation*}
\prob(|[A_t^\T A_t]_{q,q} - 1| \ge \delta_{\text{d}}) \le 2 \exp\left(\frac{-2n\delta_{\text{d}}^2}{d} \right).
\end{equation*}

Next we consider the off-diagonal entries of the diagonal blocks. There are two special cases to consider. In the special case that either $\text{mod}(p_1-q_1,d_1) \ne 0$ or $\text{mod}(p_2-q_2,d_2) \ne 0$, each of the terms in the summand in \eqref{eq:grammatrix} picks out a \emph{different} set of coefficients from $h_t$. However, the case that both $\text{mod}(p_1-q_1,d_1) = 0$ and $\text{mod}(p_2-q_2,d_2) = 0$ needs special care since there are now dependencies between the terms in the summand in \eqref{eq:grammatrix}. However, due to the special nature by which we select the coefficients, we can partition the sum into two sums (denoted $S_1$ and $S_2$ such that each of these is a sum of $n/2d$ independent terms. We can then apply the Hoeffding bound to each to yield
\begin{equation*}
\prob(|S_i| \ge \delta_{\text{o}}/2s) \le 2 \exp \left( \frac{-n\delta_\text{o}^2}{4 ds^2} \right), i = 1,2.
\end{equation*}
Then applying the union bound gives us
\begin{equation}
\label{eq:overbound}
\prob(|[A_t^\T A_t]_{p,q}| \ge \delta_{\text{o}}/s) \le 4 \exp \left( \frac{-n\delta_\text{o}^2}{4 ds^2} \right), p \ne q.
\end{equation}
Since this latter bound decays more slowly, and in the interest of simplicity, we overbound all the off-diagonal entries of the diagonal blocks using this latter expression. 

Now consider the off-diagonal blocks $A_t^\T A_{t'}$, $t \ne t'$. From independence we can show that $\expect[A_t^\T A_{t'}] = \zeros$. The entries of these matrices are sums of independent entries, and we could apply Hoeffding's bound. However, will will need to overbound with \eqref{eq:overbound} to obtain a simple expression, hence we use
\begin{equation*}
\prob(|[A_t^\T A_{t'}]_{p,q}| \ge \delta_{\text{o}}/s) \le 4 \exp \left( \frac{-n\delta_\text{o}^2}{4 ds^2} \right), t \ne t'.
\end{equation*}

Lastly, we need to perform a union bound over all the entries of the matrix. We have $nB$ diagonal entries, and exploiting symmetry, we only have $nB(nB-1)/2$ remaining entries. Hence taking $\delta_\text{d} = \delta_\text{o} = \delta_s/2$,
\begin{equation*}\begin{aligned}
&\prob(A \text{ does not satisfy } \text{RIP}(s,\delta_s)) \\
& \quad \le 2nB \exp\left(\frac{-n\delta_s^2}{2d} \right) + 4[nB(nB-1)/2] \exp \left( \frac{-n\delta_s^2}{16 ds^2} \right) \\
& \quad \le [2nB + 2nB(nB-1)] \exp \left( \frac{-n\delta_s^2}{16 ds^2} \right) \\
%& \quad = 2n^2B^2 \exp \left( \frac{-n\delta_s^2}{16 ds^2} \right)  \\
& \quad \le 2n^2B^2 \exp \left( \frac{-c_2 n}{ds^2} \right),
\end{aligned}\end{equation*}
where $c_2 \le \delta_s^2/16$. If $nB \ge 2$, this probability is less than one provided $n/d \ge c_1 \delta^{-2}s^2 \log(nB)$ for any $\delta \le \delta_s$ where $c_1 \ge 3 \delta^2/c_2$, noting that $m=n/d$ establishes the theorem.
\end{IEEEproof}

\subsection{Coarse Reconstruction using Dual-Scale Masks}
\label{app:dsm}

This section verifies \eqref{eq:lrestimate}. First, we decompose $\fstar$ into two terms,
\begin{equation}
\label{eq:fstardecomp}
\fstar_t = D^\T f_{\floor{(t-1)/B} + 1}^\text{L} + \left[\fstar_t - D^\T f_{\floor{(t-1)/B} + 1}^\text{L} \right],
\end{equation}
such that the first is based on $\fL$ and the remaining is a residual term. In the sequel we define the residual as
\begin{equation*}
\til{\fstar_t} = \fstar_t - D^\T f_{\floor{(t-1)/B} + 1}^\text{L}.
\end{equation*}
Now consider our measurements. For simplicity, assume we have these without noise. Then if we define
\begin{align*}
H_t^\text{H} &\deq \FF^{-1} \diag(\FF \hH_t) \FF,\ \text{and} \\
H_k^\text{L} &\deq \FF^{-1} \diag(\FF \hL_k) \FF,
\end{align*}
we can write the observations as
\begin{equation*}
\begin{aligned}
y_k &= \sum_{t \in T_k} A_t \fstar_t 
= S \sum_{t \in T_k} 
\left[ \alpha H_{\floor{(t-1)/B}+1}^\text{L} + \beta H_t^\text{H} \right] \fstar_t \\
&= \alpha S H_k^\text{L} \sum_{t \in T_k} D^\T f_{\floor{(t-1)/B} + 1}^\text{L} \\
&\qquad + \alpha S H_k^\text{L} \sum_{t \in T_k} \til{\fstar_t}
+ \beta S \sum_{t \in T_k}H_t^\text{H} \fstar_t \\
&= \alpha B S H_k^\text{L} D^\T \fL_k \\
&\qquad + \alpha S H_k^\text{L} \sum_{t \in T_k} \til{\fstar_t}
+ \beta S \sum_{t \in T_k} H_t^\text{H} \fstar_t,
\end{aligned}
\end{equation*}
where we used the decomposition \eqref{eq:fstardecomp} in the third equality. Turning to our coarse estimate \eqref{eq:lrestimate}, we have
\begin{equation}
\begin{aligned}
\label{eq:threeterm}
\h{\fL_k} &= \frac{1}{d} \Sigma_k^\T S H_k^\text{L} D^\T \fL_k 
+ \frac{1}{dB} \Sigma_k^\T S H_k^\text{L} \sum_{t \in T_k} \til{\fstar_t} \\
&\qquad + \frac{1}{dB}\frac{\beta}{\alpha} \Sigma_k^\T S \sum_{t \in T_k} H_t^\text{H} \fstar_t.
\end{aligned}
\end{equation}

It can be shown that $\Sigma_k^\T S H_k^\text{L} D^\T = d I$. For simplicity of notation, we show this for 1D signals; the extension to higher dimensions is straightforward. Furthermore, we drop the $k$ subscripts temporarily. We show that $SH^\text{L}D^\T = d\Sigma$, since by construction of the random phases $\Sigma^\T \Sigma = I_m$. To see this, first let $g \deq \F^{-1} \sigma$, then by construction we have
\begin{equation*}
g_k = h_{(k-1)d + 1} = \frac{1}{d} \sum_{q=1}^d h_{ (k-1)d + q}^\text{L},
\end{equation*}
and so 
\begin{equation*}
\Sigma_{kl} = g_{(k-l)_m + 1} = h_{(k-l)_m d + 1}^\text{L} = \frac{1}{d} \sum_{q=1}^d h_{ (k-l)_m d + q}^\text{L}.
\end{equation*}
Now let us examine $S H^\text{L} D^\T$:
% \begin{equation*}
% \begin{aligned}
% (S H^\text{L} D^\T)_{kl} &= \sum_{i=1}^n \sum_{j=1}^n S_{ki} H^\text{L}_{ij} D_{lj} \\
% &= \sum_{i=1}^n \sum_{j=1}^n \ones_{\{i = kd\}} h_{(i-j)_n + 1}^\text{L} \sum_{q=1}^d \ones_{\{j = (l-1)d + q\}} \\
% &= \sum_{q=1}^d h_{[ (kd) - ((l-1)d + q)]_n + 1}^\text{L} \\
% &= \sum_{q=1}^d h_{((k-l)d + d - q)_n + 1}^\text{L} \\
% &= \sum_{q=1}^d h_{((k-l)d + p - 1)_n + 1}^\text{L} \\
% &= \sum_{q=1}^d h_{(k-l)_m d + p}^\text{L} = d g_{(k-l)_m + 1},
% \end{aligned}
% \end{equation*}
\begin{equation*}
\begin{aligned}
&(S H^\text{L} D^\T)_{kl} \\
&= \sum_{i=1}^n \sum_{j=1}^n S_{ki} H^\text{L}_{ij} D_{lj} \\
&= \sum_{i=1}^n \sum_{j=1}^n \ones_{\{i = kd\}} h_{(i-j)_n + 1}^\text{L} \sum_{q=1}^d \ones_{\{j = (l-1)d + q\}} \\
&= \sum_{q=1}^d h_{[ (kd) - ((l-1)d + q)]_n + 1}^\text{L} 
= \sum_{q=1}^d h_{((k-l)d + d - q)_n + 1}^\text{L} \\
&= \sum_{q=1}^d h_{((k-l)d + p - 1)_n + 1}^\text{L} 
= \sum_{q=1}^d h_{(k-l)_m d + p}^\text{L} = d g_{(k-l)_m + 1},
\end{aligned}
\end{equation*}
and hence $SH^\text{L}D^\T = d\Sigma$ as claimed.

Because $\Sigma_k^\T S H_k^\text{L} D^\T = d I$, we have 
\begin{equation*}
\h{\fL_k} = \fL_k + \eta_k,
\end{equation*}
where $\eta_k$ is defined as the second two terms in \eqref{eq:threeterm}. Note that we expect that $\eta_k$ will be small in comparison to $\fL_k$. In particular, the first term in $\eta_k$ will be small because the energy in $\fL_k$ will be much greater than that in $\til{\fstar_t}$ for smoothly varying video sequences. The second term in $\eta_k$ will be zero-mean because $\sigma_k$ and $\hH_t$ are generated independently from zero-mean distributions.

\ifCLASSOPTIONcaptionsoff
  \newpage
\fi

\bibliographystyle{IEEEbib}
\bibliography{TIPCoded}

\begin{IEEEbiography}{Zachary T.~Harmany} received the B.S. (magna cum lade, with honors) in Electrical Engineering and B.S. (cum lade) in Physics from The Pennsylvania State University in 2006 and his Ph.D. in Electrical and Computer Engineering from Duke University in 2012. In 2010 he was a visiting researcher at The University of California, Merced. He is currently a Postdoctoral Fellow at the University of Wisconsin-Madison. His research interests include nonlinear optimization, machine learning, and signal and image processing with applications in medical imaging, astronomy, and night vision. 
\end{IEEEbiography}

\begin{IEEEbiography}{Roummel F.~Marcia}
received his B.A. in Mathematics from Columbia  University in 1995 and his Ph.D. in
Mathematics from the University of California, San Diego in 2002.  He was a Computational and Informatics in Biology and Medicine Postdoctoral Fellow in the Biochemistry Department at the University of Wisconsin-Madison and a Research Scientist in the Electrical and Computer Engineering at Duke University.  He is currently an Associate Professor of Applied Mathematics at the University of California, Merced.  His research interests include nonlinear optimization, numerical linear algebra, signal and image processing, and computational biology.
\end{IEEEbiography}

%\newpage

\begin{IEEEbiography}{Rebecca M.~Willett} (SM '02, M '05, SM '11) is an assistant professor in the Electrical and Computer Engineering Department at Duke University. She completed her PhD in Electrical and Computer Engineering at Rice University in 2005. Prof. Willett received the National Science Foundation CAREER Award in 2007, is a member of the DARPA Computer Science Study Group, and received an Air Force Office of Scientific Research Young Investigator Program award in 2010. Prof. Willett has also held visiting researcher positions at the Institute for Pure and Applied Mathematics at UCLA in 2004, the University of Wisconsin-Madison 2003-2005, the French National Institute for Research in Computer Science and Control (INRIA) in 2003, and the Applied Science Research and Development Laboratory at GE Healthcare in 2002. Her research interests include network and imaging science with applications in medical imaging, wireless sensor networks, astronomy, and social networks. %Additional information, including publications and software, are available online at http://www.ee.duke.edu/~willett/.
\end{IEEEbiography}


\vfill


\end{document}